\documentclass{mn2e}
\usepackage{epsfig,natbib2,natbibmnfix}
\usepackage{amssymb}
\usepackage{multirow}
\usepackage{multicol}
\usepackage[varg]{txfonts}
\usepackage{color}
\usepackage{rotating}
\newcommand{\cref}[1]{Chapter~\ref{#1}}

\begin{document}
%%%%%%%%%%%%%%%%%%%%%%%%%%%%%%%%%%%%%%%%%%%%%%%%%%%%%%%%%%%%%%%%%%%%%%%%%%%%%%
%% Title Details and Page Header                                            %%
%%%%%%%%%%%%%%%%%%%%%%%%%%%%%%%%%%%%%%%%%%%%%%%%%%%%%%%%%%%%%%%%%%%%%%%%%%%%%%
\title[The origin of circumstellar features in the spectra of hot DA white dwarfs]{The origin of hot white dwarf circumstellar features}

\author[N.J.~Dickinson et al.] 
{
\parbox{5in}{N.J. Dickinson$^1$, M.A. Barstow$^1$, B.Y. Welsh$^2$, M. Burleigh$^1$, J. Farihi$^1$, S. Redfield$^3$, and K. Unglaub$^4$}
\vspace{0.1in} 
 \\ $^1$ Department of Physics \& Astronomy, University of Leicester,
 Leicester, LE1 7RH, UK  
 \\ $^2$ Space Sciences Laboratory, University of California, Berkeley, California, USA
 \\ $^3$ Astronomy Department, Van Vleck Observatory, Wesleyan University, Middletown, CT 06459, USA
 \\ $^4$ Dr. Remeis-Sternwarte, Astronomisches Institut der Universit\"{a}t Erlangen-N\"{u}rnberg, Sternwartstrasse 7, 96049 Bamberg, Germany  
}

\maketitle

%%%%%%%%%%%%%%%%%%%%%%%%%%%%%%%%%%%%%%%%%%%%%%%%%%%%%%%%%%%%%%%%%%%%%%%%%%%%%%
%% Abstract, Keywords and contact details                                   %%
%%%%%%%%%%%%%%%%%%%%%%%%%%%%%%%%%%%%%%%%%%%%%%%%%%%%%%%%%%%%%%%%%%%%%%%%%%%%%%
\begin{abstract}
We have analysed a sample of 23 hot DAs to better understand the source of the circumstellar features reported in previous work. Unambiguous detections of circumstellar material are again made at eight stars. The velocities of the circumstellar material at three of the white dwarfs are coincident with the radial velocities of ISM along the sight line to the stars, suggesting that the objects may be ionising the ISM in their locality. In three further cases, the circumstellar velocities are close to the ISM velocities, indicating that these objects are either ionising the ISM, or evaporated planetesimals/material in a circumstellar disc. The circumstellar velocity at WD 1614$-$084 lies far from the ISM velocities, indicating either the ionisation of an undetected ISM component or circumstellar material. The material seen at WD\,0232+035 can be attributed to the photoionisation of material lost from its M dwarf companion. The measured column densities of the circumstellar material lie within the ionised ISM column density ranges predicted to exist in hot DA Str\"{o}mgren spheres.

\end{abstract}


\footnotetext[1]{E-mail: njd15@le.ac.uk}

%%%%%%%%%%%%%%%%%%%%%%%%%%%%%%%%%%%%%%%%%%%%%%%%%%%%%%%%%%%%%%%%%%%%%%%%%%%%%%
%% Introduction                                                             %%
%%%%%%%%%%%%%%%%%%%%%%%%%%%%%%%%%%%%%%%%%%%%%%%%%%%%%%%%%%%%%%%%%%%%%%%%%%%%%%
\section{Introduction}
\label{intro}

During the past few decades of white dwarf research, non-photospheric absorption features have been seen in far ultraviolet (FUV) white dwarf spectra, in addition to those from highly ionised photospheric material. For example, two sets of C\,{\sc iv} and Si\,{\sc iv}  absorption features are seen in the \textit{IUE} spectrum of WD\,0232+035 \citep[Feige\,24;][]{DupreeRaymond82}. This was interpreted as a set of photospheric features (with changing velocity, due to binary motion of the white dwarf) and a set of stationary features arising in an ionised gas. Si\,{\sc ii}, Si\,{\sc iii} and C\,{\sc ii} absorption features displaced by $-$12\,km\,s$^{-1}$ with respect to the photosphere are seen at WD\,1620$-$391 \citep[CD $-$38$^{\rm o}$ 10980;][]{HolbergBruhweilerAnderson95}. Given that the interstellar medium (ISM) velocity observed along the line of sight of the star does not coincide with the velocities of the circumstellar lines, these excited features are used to infer the presence of circumstellar material (to retain consistency with other work, the `circumstellar' features we discuss in this paper are the non-photospheric high ion absorption components that have been designated as circumstellar components by previous authors, e.g. \citealt{HBS98}, \citealt{HBS99}, \citealt{Bannister03}, due to the lack of similarity between the circumstellar line velocities and the photospheric/ISM velocities).

In a survey of 55 \textit{IUE} white dwarf spectra, \cite{HBS98} found that 11 stars displayed circumstellar material, of which five were DAs. All the circumstellar features were found to occupy a narrow, blueshifted velocity range (40\,-\,60\,km\,s$^{-1}$), and were attributed to stellar mass loss. Lines of sight near the objects showed no similar absorption, and unlike the circumstellar features, the velocities of the ISM absorption lines along the sight lines to the stars were both red and blueshifted \citep{HBS98,HBS99}. In a more recent survey, \cite{Bannister03} found evidence of circumstellar material at eight of the 23 DAs they examined, with two possible further detections. Proposed sources of the circumstellar features were material in the gravitational well of the stars, the ionisation of nearby ISM, stellar mass loss and material related to ancient planetary nebulae (PNe). 

Since the study of \cite{Bannister03}, research into circumstellar material at cooler white dwarfs has flourished. Given the short diffusion timescales of metals in DAs with \textit{T}$_{\rm eff}$ $<$ 25\,000\,K \cite[e.g.][]{KoesterWilken06}, an external source of polluting material is needed (the effect of radiative levitation, though not negligible, does not account for the observed metal abundances in this temperature regime; \citealt{ChayerDupuis10}, \citealt{Dupuisetal10}). Infrared studies have found evidence for dust discs around some of the cooler DAZ stars, from which metals are accreted (e.g. \citealt{Kilicetal05}, \citealt{KilicRedfield07}, \citealt{vonHippel07}, \citealt{FarihiZuckermanBecklin08}). The tidal disruption of minor planets or asteroids was first put forward as the source of this material by \cite{DebesSigurdsson02}. Subsequent studies (e.g. \citealt{Zuckermanetal03}, \citealt{Jura03}, \citealt{Kilicetal05}, \citealt{Kilicetal06}, \citealt{Jura06}, \citealt{Jura08}, \citealt{FarihiJuraZuckerman09}, \citealt{JuraFarihiZuckerman09}, \citealt{Farihietal10}) found further evidence for this. Gaseous components have been found at some white dwarf circumstellar discs (\citealt{Gansickeetal06}, \citealt{Gansickeetal07}, \citealt{Gansickeetal08}, \citealt{Melisetal11}). 

Potential evidence for circumstellar discs has also been observed at hotter white dwarfs. Four of the DAs in the WIRED survey \citep{Debesetal11} with \textit{T}$_{\rm eff}$\,$>$\,30\,000\,K have infrared excesses possibly due to dust discs. Also, an infrared excess is seen in the spectrum of the 110\,000\,K central star of the Helix nebula (WD\,2226$-$210), which could be due to a dust disc \citep{Suetal07}. In this case, the dust disc may be due to a Kuiper belt/Oort cloud analogue. In a recent survey of 71 hot white dwarfs by \cite{Chuetal11}, 35 stars were central stars of planetary nebulae (CSPN), and 20 per cent of them exhibited an infrared excess. It must be stated, however, that the precise origin of these infrared excesses is not yet fully understood, and may indeed be related to the nebulae or binary companions seen at some of these hot stars. Only five to six percent of the non-CSPN had similar infrared excesses.

Some previous studies looked to white dwarf Str\"{o}mgren spheres to explain observations of the ISM. \cite{DupreeRaymond83} used Str\"{o}mgren spheres to explain the high ion absorption features seen in the \textit{IUE} spectra of WD\,0232+035 and WD\,0501+527 (G191$-$B2B), though these features are now known to  be due to photospheric metals. \cite{TatTerzian99} found that the UV ionisation of the ISM within 20\,pc of the Sun could be due to hot white dwarf Str\"{o}mgren spheres. 121 stars were studied, of which 24 had estimated Str\"{o}mgren sphere radii (\textit{r}$_{S}$)\,$>$\,0.5\,pc. However, the influence of B stars within 100\,pc \citep{Vallerga98} was not accounted for, nor was the presence of the rarefied tunnel towards $\beta$ CMa \citep{Welsh91} and the local chimney \citep{Welshetal99}. \cite{Welshetal10b} observed a series of cell-like cavity structures in their 3-D maps of Na\,{\sc i} and Ca\,{\sc ii} in the LISM, attributed to nearby B stars and some hot white dwarfs.

\cite{Lallementetal11} cite the evaporation and ionisation of circumstellar material by  hot white dwarfs as an explanation of the circumstellar features observed in their sample. An important point highlighted by \cite{Lallementetal11} was that if the O\,{\sc vi} absorption studied by \cite{SL06} and \cite{Barstowetal10} is in fact circumstellar, then erroneous interpretations of the physical state of the LISM  may have been made, which could explain the unclear view of the morphology of local hot interstellar gas. \cite{Welshetal10} and \cite{Welshetal10b} also found evidence for the ionisation of the LISM by hot white dwarfs and B stars where ISM hot/cold interfaces have previously been used to explain the observed high ions. This shows a clear need to understand the true nature of the high ions seen in hot DA spectra, to avoid misinterpretation of non-photospheric absorption features.

Since the study of \cite{Bannister03}, improvements have been made in both the measuring of ISM/circumstellar absorption features (section \ref{measuring})  and in the understanding of white dwarf circumstellar environments. Here, we re-examine the sample of \cite{Bannister03}, to better understand the origin of the observed circumstellar absorption features.

%%%%%%%%%%%%%%%%%%%%%%%%%%%%%%%%%%%%%%%%%%%%%%%%%%%%%%%%%%%%%%%%%%%%%%%%%%%%%%
%% Measuring circumstellar absorption components                            %%
%%%%%%%%%%%%%%%%%%%%%%%%%%%%%%%%%%%%%%%%%%%%%%%%%%%%%%%%%%%%%%%%%%%%%%%%%%%%%%
\section{Measuring circumstellar absorption components}
\label{measuring}

The stellar parameters and observation details for the DAs studied here are presented in table \ref{table:tloggobs}. The high ion absorption lines examined are listed in table \ref{table:ionwavelen}. When high ion absorption features with more than one component were observed in the \textit{STIS}/\textit{IUE} spectra, the O\,{\sc vi} features in the \textit{FUSE} (\textit{R} = 20\,000) spectrum of the star were modelled, where data was available and the features were present. 
The lines were fit using the method outlined in the work of \cite{WelshLallement05} and \cite{WelshLallement10}, and a brief summary is given here. The continuum around the absorption features was measured using a third order polynomial, and the line profiles were then fit with Gaussian absorption components. The (heliocentric) velocities of the model components were allowed to vary through a $\chi^{2}$ minimisation technique to obtain a best fit. The \textit{b} values of the lines were measured in a similar way. Theoretical absorption line profiles for all circumstellar and interstellar components were calculated using the measured \textit{b} values and line oscillator strengths, and were used to measure a column density in every case. Non-photospheric, high ion absorption components were added to the model in addition to the photospheric component when the fit was improved in a statistically significant way, i.e. if the change in the absolute $\chi^{2}$ was greater than 11.1 with the additional absorbing component (at which point the probability of this improvement in  $\chi^{2}$ being random is less than 1\%; \citealt{Vallergaetal93}). This self-consistent modelling of all non-photospheric absorption components offers significant improvement over the technique used by \cite{Bannister03}, where a curve of growth was used to obtain a circumstellar column density for a series of discrete \textit{b} values in most circumstellar C\,{\sc iv} and Si\,{\sc iv} features, since column densities are measured for all non-photospheric absorption features. Reliable errors were produced for unsaturated absorption lines. Since the oscillator strength was used in the modelling of all absorption lines, the technique utilised by \cite{Bannister03} of coadding doublet components in velocity space to reduce the signal to noise and reveal hidden circumstellar components cannot be made use of here.

\begin{table*}
\caption{The stellar parameters and observation information for the white dwarfs observed.}
\begin{tabular}{c c c c c c c c c}
\hline
WD             & Alt.name        & \textit{T}$_{\rm eff}$ (K)$^a$ & log\,\textit{g}$^a$ & \textit{L/L}$_{\odot}^b$ & \textit{D}(pc)$^b$ & Data source [Mode]                      & Resolving Power \\
\hline   
0050$-$335     & GD\,659         & 35\,660$\pm$135                & 7.93$\pm$0.03        & 0.24                    & 53                 & \textit{STIS}[E140M]; \textit{IUE}[SWP] & 40\,000; 20\,000\\
0232+035       & Feige\,24       & 60\,487$\pm$1\,100             & 7.50$\pm$0.06        & 5.86                    & 78                 & \textit{STIS}[E140M]                    & 40\,000\\
0455$-$282     & REJ\,0457$-$281 & 50\,960$\pm$1\,070             & 7.93$\pm$0.08        & 1.85                    & 108                & \textit{IUE}[SWP]                       & 20\,000\\
0501+527       & G191$-$B2B      & 52\,500$\pm$900                & 7.53$\pm$0.09        & 3.16                    & 50                 & \textit{STIS}[E140M]                    & 40\,000\\
0556$-$375     & REJ\,0558$-$373 & 59\,508$\pm$2\,200             & 7.70$\pm$0.09        & 4.61                    & 295                & \textit{STIS}[E140M]                    & 40\,000\\
0621$-$376     & REJ\,0623$-$371 & 58\,200$\pm$1\,800             & 7.14$\pm$0.11        & 11.69                   & 97                 & \textit{IUE}[SWP]                       & 20\,000\\
0939+262       & Ton\,021        & 69\,711$\pm$530                & 7.47$\pm$0.05        & 10.19                   & 217                &\textit{STIS}[E140M]                     & 40\,000\\
0948+534       & PG\,0948+534    & 110\,000$\pm$2\,500            & 7.58$\pm$0.06        &                         & 193.8              & \textit{STIS}[E140M]                    & 40\,000\\
1029+537       & REJ\,1032+532   & 44\,350$\pm$715                & 7.81$\pm$0.08        & 1.03                    & 127                & \textit{STIS}[E140M]                    & 40\,000\\
1057+719       & PG\,1057+719    & 39\,770$\pm$615                & 7.90$\pm$0.10        & 0.64                    & 411                & \textit{GHRS}[G160M]                    & 22\,000\\
1123+189       & PG\,1123+189    & 54\,574$\pm$900                & 7.48$\pm$0.08        & 2.75                    & 147                & \textit{STIS}[E140H]                    & 100\,000\\
1314+493       & HZ\,43          & 50\,370$\pm$780                & 7.85$\pm$0.07        & 1.49                    & 71                 & \textit{IUE}[SWP]                       & 20\,000\\
1254+223       & GD\,153         & 39\,290$\pm$340                & 7.77$\pm$0.05        & 0.56                    & 73                 & \textit{IUE}[SWP]                       & 20\,000\\
1337+705       & EG\,102         & 22\,090$\pm$85                 & 8.05$\pm$0.01        & 0.03                    & 25                 & \textit{IUE}[SWP]                       & 20\,000\\
1611$-$084     & REJ\,1614$-$085 & 38\,840$\pm$480                & 7.92$\pm$0.07        & 0.43                    & 86                 & \textit{GHRS}[G160M]                    & 22\,000\\
1738+669       & REJ\,1738+665   & 66\,760$\pm$1\,230             & 7.77$\pm$0.10        & 12.59                   & 243                & \textit{STIS}[E140M]                    & 40\,000\\
2023+246       & Wolf 1346       & 19\,150$\pm$30                 & 7.91$\pm$0.01        & 0.03                    & 14                 & \textit{IUE}[SWP]                       & 20\,000\\
2111+498       & GD\,394         & 39\,290$\pm$360                & 7.89$\pm$0.05        & 0.45                    & 57                 & \textit{IUE}[SWP]; \textit{GHRS}[G160M] & 20\,000; 22\,000\\
2152$-$548     & REJ\,2156$-$546 & 45\,500$\pm$1\,085             & 7.86$\pm$0.10        & 1.06                    & 129                & \textit{STIS}[E140M]                    & 40\,000\\
2211$-$495     & REJ\,2214$-$492 & 61\,613$\pm$2\,300             & 7.29$\pm$0.11        & 9.38                    & 69                 & \textit{IUE}[SWP]                       & 20\,000\\
2218+706       & WD\,2218+706    & 58\,582$\pm$3\,600             & 7.05$\pm$0.12        & 9.64                    & 436                & \textit{STIS}[E140M]                    & 40\,000\\
2309+105       & GD\,246         & 51\,308$\pm$850                & 7.91$\pm$0.07        & 2.23                    & 72                 & \textit{STIS}[E140M]; \textit{IUE}[SWP] & 40\,000; 20\,000\\
2331$-$475     & REJ\,2334$-$471 & 53\,205$\pm$1\,300             & 7.67$\pm$0.10        & 2.88                    & 104                & \textit{IUE}[SWP]                       & 20\,000\\
\hline  &               
\end{tabular}
\label{table:tloggobs}
\\$^a$from \cite{Barstowetal03};$^b$from \cite{Bannister03} and references therein
\end{table*}

The photospheric velocity ($v$$_{\rm phot}$) was obtained by calculating the mean of the velocities of the photospheric components in each of the absorption features.  The averaged circumstellar velocity ($v$$_{\rm CS}$) was obtained in a similar way. The ISM was characterised by measuring the Si\,{\sc ii} (1260.422\,\AA\, 1304.370\,\AA) and S\,{\sc ii} (1259.52\,\AA) lines, since they were often unsaturated and enabled more accurate velocity measurements. When additional measurements were required, for example when either Si\,{\sc ii} or S\,{\sc ii} were not present along the sight line, other lines such as O\,{\sc i} (1302.168\,\AA) or Fe\,{\sc ii} (1608.451\,\AA) were used. The ISM components were denoted in order of equivalent width, i.e. the component with the largest equivalent width was designated the primary ISM component (with $v_{\rm ISM,pri}$), the second largest was the secondary component (with $v_{\rm ISM,sec}$), etc. Mean values of $v_{\rm ISM,pri}$, $v_{\rm ISM,sec}$ and $v_{\rm ISM,ter}$ were calculated, to allow the value of $v_{\rm CS}$ to be compared to each of the ISM components along the sight line in each case. All errors were combined quadratically.

\begin{table}
\caption{The laboratory wavelengths of the high ion absorption features studied here.}
\begin{tabular}{c c}
\hline
Ion          & Lab. wavelength (\AA)\\
\hline
C\,{\sc iv}  & 1548.187, 1550.772\\
N\,{\sc v}   & 1238.821, 1242.804\\
O\,{\sc v}  & 1371.296\\
O\,{\sc vi}  & 1031.912, 1037.613\\
Si\,{\sc iv} & 1393.755, 1402.770\\
\hline
\end{tabular}
\label{table:ionwavelen}
\end{table}

%%%%%%%%%%%%%%%%%%%%%%%%%%%%%%%%%%%%%%%%%%%%%%%%%%%%%%%%%%%%%%%%%%%%%%%%%%%%%%
%% Results                                                                  %%
%%%%%%%%%%%%%%%%%%%%%%%%%%%%%%%%%%%%%%%%%%%%%%%%%%%%%%%%%%%%%%%%%%%%%%%%%%%%%%
\section{Results}
\label{results}

Eight white dwarfs show unambiguous signs of high ion, non-photospheric material in their \textit{HST/IUE} spectra. Of these stars, a clear circumstellar O\,{\sc vi} detection is made in the \textit{FUSE} spectrum of one star (WD\,1738+669). The traces of circumstellar C\,{\sc iv} found at WD\,0050$-$335 and WD\,2152$-$548 by \citealt{Bannister03} are not confirmed here. Table \ref{table:vels} details the measured absorption component velocities and the circumstellar velocity shifts with respect to $v$$_{\rm phot}$. The ISM components (primary, `pri'; secondary, `sec'; tertiary `ter') are ordered from largest to smallest equivalent width. Given the proximity of some of the stars in this sample, projected velocities were calculated for the LISM clouds traversed by the line of sight toward each white dwarf\footnote{the cloud identification and velocity projections were performed using the online tool available at http://lism.wesleyan.edu/LISMdynamics.html}, using the LISM morphology maps of \cite{RedfieldLinsky08}. Indeed, some of the lower ionisation ISM components seen along the sight lines to the objects studied here can be expected to originate in the LISM (these are discussed for each object in section \ref{appendix}). The clouds and their predicted velocities are detailed in the sixth column of table \ref{table:vels}. Gravitational redshifts for each DA are shown in the seventh column. Table \ref{table:cols} details, for each star, which of the high ions display circumstellar components, and gives their column densities. Detailed results for each object are presented in section \ref{appendix}.

\begin{centering}
\begin{table*}
\caption[]{All measured velocities, circumstellar velocity shifts ($v$$_{\rm CS shift}$), predicted LISM cloud velocities ($v_{\rm LISM,pred}$)  and gravitational redshifts ($v_{\rm grav}$) for each white dwarf. All velocities are heliocentric and are expressed in km\,s$^{-1}$.}
\scriptsize{
\begin{tabular}{c c c c c c c}
\hline
WD           & $v_{\rm phot}$                         & $v_{\rm CS}$      & $v_{\rm CS shift}$                           & $v_{\rm ISM(pri,sec,ter)}$                       & $v_{\rm LISM,pred}$ (cloud name)$^\ast$                     & $v_{\rm grav}$\\
\hline
0050$-$335 & 34.34$\pm$0.38                         &                   &                                              & 6.7$\pm$0.3                                      & 4.56$\pm$1.36 (Local Interstellar Cloud, LIC)   & 28.21\\
0232+035   & 30.11$\pm$0.52$^a$,128.23$\pm$0.31$^b$ & 7.4$\pm$0.34      & $-$22.19$\pm$0.68$^a$,-120.63$\pm$0.40$^b$       & 2.85$\pm$0.34,17$\pm$1.3                         & 18.1$\pm$1.13 (LIC)                             & 15.70\\
0455$-$282 & 79.28$\pm$1.79                         & 18.8$\pm$3.47     & $-$60.48$\pm$3.90                               & 12.1$\pm$1.5                                     & 12.56$\pm$1.03 (Blue)                           & 28.96\\
0501+527   & 24.51$\pm$0.16                         & 8.9$\pm$0.07      & $-$15.61$\pm$0.17                               & 8.15$\pm$0.18,19.3$\pm$0.03                      & 19.1$\pm$1.1 (LIC), 9.35$\pm$1.32 (Hyades)      & 16.07\\  
0556$-$375 & 25.37$\pm$2.03                         & 10.2$\pm$1.07     & $-$15.17$\pm$2.30                               & 7.8$\pm$1                                        & 11.36$\pm$0.95 (Blue)                           & 21.07\\
0621$-$376 & 39.44$\pm$0.25                         &                   &                                              & 15.8$\pm$0.4                                     & 11.09$\pm$0.93 (Blue)                           & 9.36\\
0939+262   & 36.5$\pm$0.47                          & 9.38$\pm$6.6      & $-$27.12$\pm$6.61                               & $-$2.1$\pm$.0.2                                 & 10.81$\pm$1.29 (LIC)                            & 15.40\\
0948+534   & $-$17.09$\pm$1.73                      &                   &                                              & $-$18.45$\pm$0.42, $-$1.6$\pm$0.63, 22.6$\pm$0.8 & 10.07$\pm$1.31 (LIC)                            & 19.67\\
1029+537   & 37.98$\pm$0.21                         &                   &                                              & 0.95$\pm$0.79                                    & 7.72$\pm$1.33 (LIC)                             & 24.00\\
1057+719   &                                        &                   &                                              & $-$0.2$\pm$1                                     & 6.64$\pm$1.35 (LIC)                             & 27.18\\
1123+189   &                                        &                   &                                              & $-$4.75$\pm$3.18                                 & 3.03$\pm$0.79 (Leo)                             & 14.98\\
1314+493   &                                        &                   &                                              & $-$6.6$\pm$0.1                                   & $-$6.15$\pm$0.74 (NGP)                          & 25.77\\
1254+223   &                                        &                   &                                              & $-$15.4$\pm$1.8                                  & $-$5.52$\pm$0.74 (NGP)                          & 22.36\\
1337+705   &                                        &                   &                                              & $-$1.5$\pm$1.8                                   & 1.59$\pm$1.38 (LIC)                             & 32.84\\
1611$-$084 & $-$40.76$\pm$3.56                      & $-$66.67$\pm$2.05 & $-$25.90$\pm$4.11                               & $-$34.7$\pm$1.5                                  & $-$29.26$\pm$1.12 (G)                           & 27.95\\
1738+669   & 30.17$\pm$1.49                         & $-$18.36$\pm$4.23 & $-$48.53$\pm$4.49                               & $-$20.0$\pm$0.3                                  & $-$2.91$\pm$1.37 (LIC)                          & 23.65\\
2023+246   &                                        &                   &                                              & $-$16.3$\pm$1.7, 18.3$\pm$2.5                    &                                                 & 26.41\\
2111+498   & 29.3$\pm$1.66                          &                   &                                              & $-$7.6$\pm$1.3                                   & $-$2.35$\pm$1.38 (LIC)                          & 26.77\\
2152$-$548 & $-$14.94$\pm$0.46                      &                   &                                              & $-$9.2$\pm$0.53                                  & $-$9.73$\pm$1.31 (LIC)                          & 25.90\\
2211$-$495 & 32.33$\pm$1.37                         &                   &                                              & $-$1.1$\pm$0.4                                   & $-$8.8$\pm$1.32 (LIC), 9.93$\pm$0.6 (Dor)       & 11.43\\
2218+706   & $-$40.04$\pm$1.11                      &$ -$17.8$\pm$1.05  & 22.24$\pm$1.52                            & $-$15.3$\pm$2.64, $-$1.2$\pm$4.01                & 4.47$\pm$1.37 (LIC)                             & 8.04\\
2309+105   & $-$13.45$\pm$0.13                      &                   &                                              & $-$8.2$\pm$0.70                                  &                                                 & 28.17\\
2331$-$475 & 38.88$\pm$0.72                         &                   &                                              & 14.3$\pm$0.7                                     & $-$3.41$\pm$1.37 (LIC)                          & 19.88\\ 
\hline
\end{tabular}
}
\label{table:vels}
\\$^\ast$from \cite{RedfieldLinsky08};$^a$from binary phase 0.24;$^b$from binary phase 0.74 
\end{table*}
\end{centering}

\begin{table*}
\caption{The stars with unambiguous circumstellar detections, the identified species and the measured column densities.}
\begin{tabular}{c c c}
\hline
WD             & Species                                & Column density (10$^{12}$cm$^{-2}$)\\
\hline
0232+035     & C\,{\sc iv}                            & 28.0$\pm$1.3\\
0455$-$282   & C\,{\sc iv}, N\,{\sc v}, Si\,{\sc iv}  & 33.2$\pm$6.6, 4.17$\pm$0.83, 3.71$\pm$0.74\\
0501+527     & C\,{\sc iv}                            & 104$\pm$10\\  
0556$-$375   & C\,{\sc iv}                            & 46.8$\pm$6.2\\
0939+262     & C\,{\sc iv}, Si\,{\sc iv}              & 8.05$\pm$0.21, 1.81$\pm$0.36\\
1611$-$084   & C\,{\sc iv}, Si\,{\sc iv}              & 9.77$\pm$1.9, 3.53$\pm$0.71\\
1738+669     & C\,{\sc iv}, O\,{\sc v},O\,{\sc vi}, Si\,{\sc iv} & 55.5$\pm$6.1, 2.35$\pm$0.29, 7.04$\pm$6.73, 3.77$\pm$0.1\\
2218+706     & C\,{\sc iv}, Si\,{\sc iv}              & 119$\pm$17, 7.34$\pm$0.06\\
\hline
\end{tabular}
\label{table:cols}
\end{table*}

%%%%%%%%%%%%%%%%%%%%%%%%%%%%%%%%%%%%%%%%%%%%%%%%%%%%%%%%%%%%%%%%%%%%%%%%%%%%%%
%% Discussion                                                               %%
%%%%%%%%%%%%%%%%%%%%%%%%%%%%%%%%%%%%%%%%%%%%%%%%%%%%%%%%%%%%%%%%%%%%%%%%%%%%%%
\section{Discussion}
\label{discussion}

With a few exceptions, the velocities found in this work are broadly consistent with those reported by \cite{Bannister03}. Full modelling of the circumstellar components has, for the first time, allowed column densities to be measured for all circumstellar species at all stars. A hint of a second component in the high ion absorption lines of  WD\,0948+534 is seen, though a single component fit is statistically preferred. In this section, possible origins of this circumstellar material are explored, including circumstellar discs (section \ref{circumstellar discs}), ionised ISM (section \ref{Ionised ISM}), stellar mass loss (section \ref{Mass loss}) and ancient planetary nebulae (section \ref{Ancient planetary nebulae}). A summary is presented in section \ref{Summary}.

\subsection{Circumstellar discs}
\label{circumstellar discs}

The circumstellar discs observed at cooler, metal polluted white dwarfs (e.g. \citealt{Farihietal10}) reside within a few tens of stellar radii from the star. If the non-photospheric lines observed here are due to analogous circumstellar discs, one would expect such discs to be sublimated so close to the host star \citep[e.g. ][]{vonHippel07}. In a search for gaseous disc components similar to those seen at some cooler white dwarfs, which included some of the stars surveyed here, \cite{Burleighetal10} and \cite{Burleighetal11} found no evidence of Ca {\sc ii}, Si {\sc ii} or Fe {\sc ii} emission. Furthermore, the surveys of \cite{Mullallyetal07} and \cite{Chuetal11} did not find infrared excesses at any of the stars in this sample (though an infrared excess is seen at WD\,2218+706, possibly due to the PN around the star).    

However, the lack of metal line emission or infrared excess detections should not be taken as strong evidence for the absence of circumstellar discs. \cite{ChayerDupuis10} and \cite{Dupuisetal10} found that, while radiative levitation has some effect below 25\,000\,K, accretion must be ongoing at WD\,0310$-$688, WD\,0612+177, WD\,1337+705, WD\,1620$-$391 (which also has circumstellar absorption features in its UV spectrum; \citealt{HolbergBruhweilerAnderson95}) and WD\,2032+248, to explain the photospheric metal abundances of the stars. A similar scenario exists for WD\,2111+498, with its Si and Fe overabundance (\citealt{Holbergetal97}, \citealt{Dupuisetal00}, \citealt{Chayeretal00}, \citealt{Vennesetal06}). Coupled with the fact that of the cooler DAZs which have photospheric metal abundances that must result from the circumstellar pollution, no more than 20\% show dust disks in the infrared (\citealt{FarihiJuraZuckerman09}, \citealt{Juraetal07}), it is possible that the stars with circumstellar features studied here have so far undetected circumstellar discs about them.

Table \ref{table:vels} shows that of the eight white dwarfs with non-photospheric absorption lines, two (WD\,0501+527 and WD\,1611$-$084) have circumstellar velocity shifts with magnitudes comparable to the $v_{\rm grav}$ of the star, implying that the observed material may be within a few tens of stellar radii of the star. \citealt{Lallementetal11} suggest that the circumstellar high ion absorption detected at the three DAs in their study may be due to the evaporation and ionisation of circumstellar planetesimals. The ionisation of material in a circumstellar disc about the star may also provide such circumstellar absorption. Such an interpretation may explain the observations here. Like the study of \cite{Lallementetal11}, the values of  $v_{\rm CS}$ at WD\,0455$-$282, WD\,0556$-$375 and WD\,0939+262 are separate from the ISM components detected along the sight lines to the stars. However, given how close the $v_{\rm CS}$ of these stars are to the $v_{\rm LISM,pred}$ values, it is not immediately clear that the material ionised by these stars is not in any way associated with the ISM, despite their large distance relative to the proximity of the LISM clouds (this is discussed in detail in the following section). The $v_{\rm CS}$ seen in the spectrum of WD\,1611$-$084 is well separated from both detected ISM and predicted LISM, again indicating that the circumstellar material may reside in a disc about the star. The precise behaviour of a circumstellar disc or planetesimals so close to the intense UV radiation field of such hot stars is not yet fully understood. A model that produces quantitative, testable predictions for circumstellar discs at such hot white dwarfs would be useful. This is, however, beyond the scope of this paper and would make an interesting further study.

\subsection{Ionised ISM}
\label{Ionised ISM}

Another possible source of the observed circumstellar features is the ionisation of the ISM. \cite{IndebetouwShull04a} outlined possible reasons for such high ions to be present in the ISM, including the evaporation of ISM cloudlets, planar conduction fronts, cooling Galactic fountain material, hot gas in stellar wind and supernova bubbles, turbulent mixing layers and white dwarf Str\"{o}mgren spheres. If the non-photospheric high ion absorption features are due to ISM processes, then one would expect to see similar non-photospheric high ion absorption features in the spectra of other stars along lines of sight near those of the white dwarfs with circumstellar absorption (except for the Str\"{o}mgren sphere model, where the high ions would be local to the star). An inspection, of \textit{IUE} and HST \textit{STIS}, \textit{GHRS} and \textit{COS} data up to five degrees from each object shows no evidence of similar non-photospheric absorption due to C\,{\sc iv}, N\,{\sc v}, O\,{\sc v} or Si\,{\sc iv} in the UV spectra along such sight lines, consistent with the result of \cite{HBS98}. This indicates that if the circumstellar features are in the ISM, they are local to the white dwarfs and not due to wider ISM processes.

\begin{figure*}
\includegraphics[angle=90, width=\textwidth]{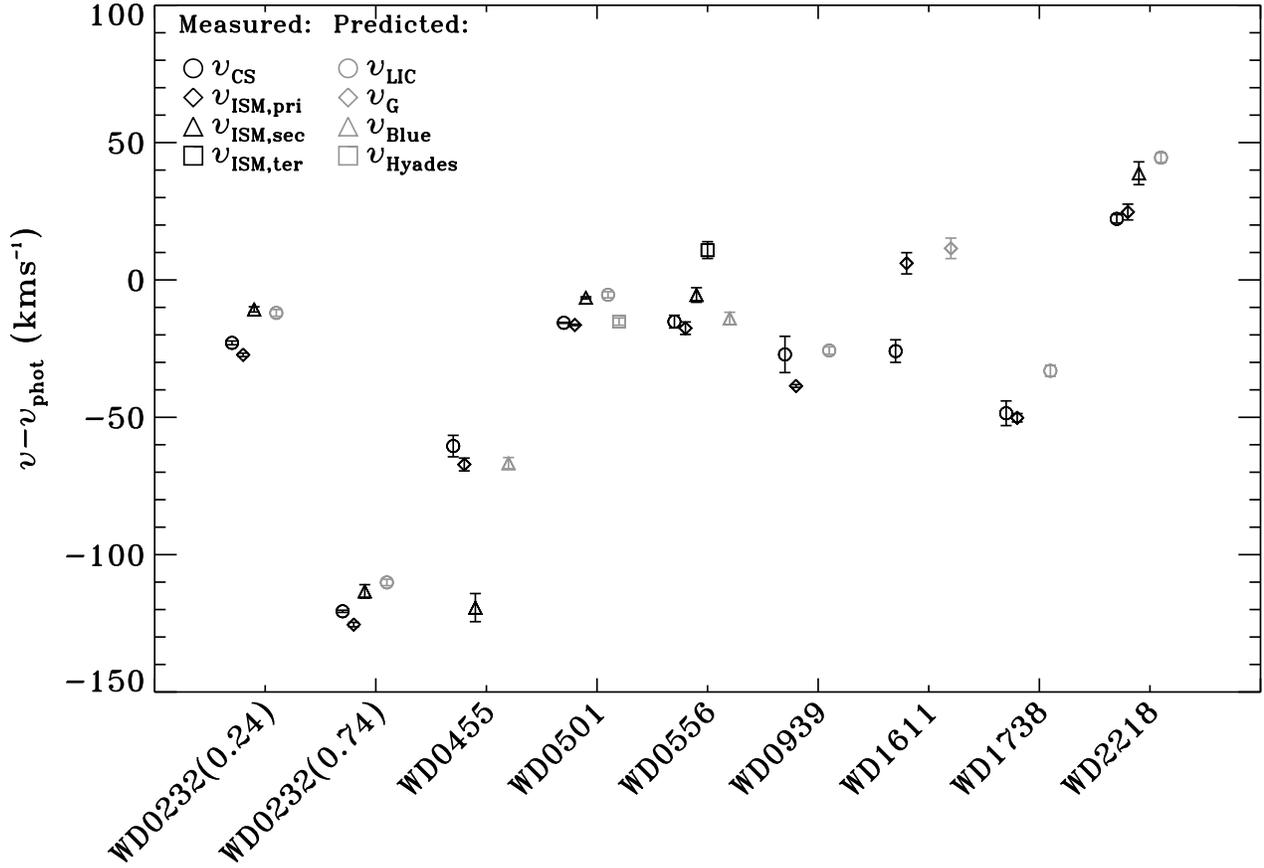}
  \caption{A plot of the shifts in $v$$_{\rm CS}$, the measured ISM component velocities ($v$$_{\rm ISMpri,sec,ter}$) and the predicted LISM cloud velocities ($v$$_{\rm LIC,G,Blue,Hyades}$), for the white dwarfs with circumstellar material. The measured ISM components are plotted in black and the predicted LISM are plotted in grey. In some cases the error bars are smaller than the plot symbols, though the plot symbols are open to allow the error bars to be seen. Two entries are present for WD\,0232+035, one for each binary phase (denoted in brackets).}
  \label{fig:vphot_fin}
\end{figure*}

Since the study of \cite{Bannister03}, improvements have been made in the mapping of the LISM. Using the maps of \cite{RedfieldLinsky08}, the LISM cloud(s) traversed by the sight line to each white dwarf  was identified and the projected velocity was calculated (table \ref{table:vels}, column six), allowing both a comparison of the detected ISM components to the predicted LISM velocities and an examination of other possible undetected ISM velocities along the sight lines to the objects. Using the measured velocities detailed in table \ref{table:vels}, one can see that in many cases $v$$_{\rm CS}$ is coincident or very close to one of the predicted $v$$_{\rm LISM,pred}$ components and/or one of the detected ISM components. Subtracting $v$$_{\rm phot}$ from these velocities (i.e. computing the velocity shift) allows an easy comparison of $v$$_{\rm CS}$ to the measured and predicted interstellar components (figure \ref{fig:vphot_fin}).

The interaction of WD\,0501+527 (at a distance of 50\,pc) with the Hyades cloud has previously been invoked to explain observations of the ISM along the sight line to the star. \cite{RedfieldFalcon08} state that the Hyades cloud is closer to WD\,0501+527 than the LIC, and the cloud has an enhanced electron density ($\sim$0.5\,cm$^{-3}$) along the sight line. This is attributed to the photoionisation of the Hyades cloud by WD\,0501+527. However, though \cite{RedfieldLinsky08} noted the common dynamical and spatial properties of both the high and low ion absorption features at $\sim$8.6 km\,s$^{-1}$, they found that the metal depletion along the line of sight to the DA is not consistent with other sights line through the Hyades cloud. Indeed, the high weighted standard deviation of metal depletion for the Hyades cloud identifies this sight line as anomalous. This may be evidence for multiple, unresolved absorbing components with similar velocities being detected along the sight line to this star, of which one is close enough to the DA to become highly ionised and produce the observed circumstellar absorption. Such an effect has been seen in other studies of the ISM, where higher resolution observations along a given sight line have resolved multiple, blended ISM components where only one was seen in data of lower resolution (e.g. \citealt{Welshetal10b}). \cite{RedfieldFalcon08} found that the LIC does not extend beyond 13 pc along the sight line to WD\,0501+527, constraining the distance to the secondary ISM component, which has a velocity in keeping with the projected LIC velocity.

Similarly, the values of $v$$_{\rm CS}$ and $v$$_{\rm ISM,pri}$  at WD\,1738+669 and WD\,2218+706 line up well with each other, indicating that an interstellar cloud near each of the stars may be being ionised. Given the distances to these stars (243 and 436\,pc), the ionisation of the LISM cannot be occurring. Indeed, the projected LIC velocities along these lines of sight are well separated from $v$$_{\rm CS}$ and $v$$_{\rm ISM}$ ($v$$_{\rm ISM,pri}$ at WD\,2218+706). The secondary component to the ISM along the sight line to WD\,2218+706 does, within error, match the predicted LIC velocity. However, the error on $v$$_{\rm ISM,sec}$ is rather large, and the association of these absorbing components on the grounds of velocity alone is tentative, and should be treated with caution.

At WD\,0455$-$282, $v_{\rm CS}$ (18.8$\pm$3.47\,km\,s$^{-1}$) is separate from both $v$$_{\rm ISM}$ (12.1$\pm$1.5\,km\,s$^{-1}$) and $v$$_{\rm LISM,pred}$ (12.56$\pm$1.03\,km\,s$^{-1}$, in this case due to the Blue cloud). Given that these velocities are quite close to one another within errors (since the LISM and associated ISM cloudlet velocities along a particular sight line occupy a narrow range; \citealt{Welshetal10b}), this could be symptomatic of the blending of ISM components at the resolution of the data here. At a distance of 108\,pc, this star may be photoionising a nearby ISM cloud that does not constitute part of the LISM, but has a velocity similar to that of the LISM. Similarly, the $v_{\rm CS}$ detection at WD\,0939+262 (9.38$\pm$6.6\,km\,s$^{-1}$) is far from $v$$_{\rm ISM}$ ($-$2.1$\pm$0.2\,km\,s$^{-1}$), while it is close to the projected LIC velocity (10.81$\pm$1.29\,km\,s$^{-1}$). Given the distance to this star (217\,pc), the ionisation of the LIC cannot be occurring. It may again be the case that a `LIC like' ISM cloud, close to the white dwarf, is being photoionised. The $v_{\rm CS}$, $v_{\rm ISM}$ and $v$$_{\rm LISM,pred}$ (10.2$\pm$1.07, 7.8$\pm$1.0 and 11.36$\pm$0.95 km\,s$^{-1}$) for WD\,0556$-$375, along with the distance to the star (295\,pc), suggest a similar situation at this object. Alternatively, as discussed in section \ref{circumstellar discs}, the high ion absorption at these three DAs may be associated with the ionisation of a circumstellar disc/planetesimals by the white dwarf, as suggested by \cite{Lallementetal11}. In addition to there being no similar non-photospheric high ion absorption along sight lines near to those of the stars observed here, the Na {\sc i} ISM map constructed by \cite{Welshetal10b} shows no nearby ISM along sight lines near these three stars at $v_{\rm CS}$ (though no data was present within 10 degrees of WD\,0939+262).

The $v$$_{\rm CS}$ of the material at WD\,0232+035 matches neither the detected ISM components nor the projected velocity of the LIC. At WD\,1611$-$084, $v$$_{\rm CS}$ is far from both the predicted and measured ISM velocities. This suggests that, with the exception of WD\,0232+035 and WD\,1611$-$084, the hot white dwarfs may be ionising the ISM in their locality. However, given the uncertainty discussed above, this should be treated with caution until higher resolution studies of the ISM can be made along these sight lines, to fully characterise any currently unresolved absorbing components.

As well as measuring the velocities of the absorbing components, column density data was also obtained (table \ref{table:cols}). \cite{IndebetouwShull04a} collated a table of predicted Si\,{\sc iv}, C\,{\sc iv}, N\,{\sc v} and O\,{\sc vi} column densities for the models in their study (table \ref{table:ISMcols}). Using the column densities measured here, only two models have a predicted column density range comparable to those observed, the  4 M$_{\odot}$ cooling fountain and the white dwarf Str\"{o}mgren sphere. The predicted column densities of the 4 M$_{\odot}$ cooling fountain model only cover a small range of the observed column densities, making white dwarf Str\"{o}mgren sphere a reasonable choice of model.

\begin{table*}
\caption{The Li-like column densities for a range of ISM models \cite[table 1,][]{IndebetouwShull04a}. All column densities are expressed in units of 10$^{12}$\,cm$^{-2}$.}
\begin{tabular}{l c c c c}
\hline
Model                                          & Si\,{\sc iv}   & C\,{\sc iv}   & N\,{\sc v}     & O\,{\sc vi}\\
\hline
\multirow{2}{*}{Evaporating cloudlet$^{a}$}    &  ...           &   1.2 - 1.5   &  0.5 - 0.6     &   9 - 12\\ 
                                               &  0.10 - 0.14   &   2.7 - 3.8   &  1.0 - 1.2     &   12 - 14\\
\multirow{2}{*}{Planar conduction front$^{a}$} &  0.10 - 0.16   &   1.6 - 3.2   &  0.6 - 1.0     &   8 - 10\\ 
                                               &  0.029 - 0.097 &   2.7 - 3.8   &  1.0 - 1.2     &   12 - 14\\
Stellar wind bubble                            &  0.21 - 0.25   &   3.3 - 4.0   &  1.3 - 1.6     &   21 - 25\\
\multirow{2}{*}{SNR bubble$^a$}                &  0.4 - 0.6     &   6.3 - 10    &  3.2 - 5.0     &   40 - 79\\ 
                                               &  $\sim$0.52    &   $\sim$7.8   &  $\sim$3.6     &   $\sim$47\\
Halo SNR bubble                                &  ...           &   8 - 15      &  3.4 - 7.9     &   35 - 150\\
4 M$_{\odot}$ cooling fountain                 &  3.3 - 6.4     &   43 - 79     &  28 - 36       &   580 - 600\\
40\,pc cooling cloud                           &  $\sim$25      &   $\sim$50    &  $\sim$13      &   $\sim$200\\
Turbulent mixing layer                         &  0.0010 - 0.47 &   0.025 - 6.8 &  0.0022 - 0.32 &  0.017 - 0.81\\
White dwarfs (Str\"{o}mgren sphere)$^{b}$      &  1.4 - 4.4     &   25 - 77     &  3.7 - 12      &   5.6 - 20\\
\hline
\end{tabular}
\label{table:ISMcols}
\\$^a$The different rows correspond to different models \citep[references contained within ][]{IndebetouwShull04a}; $^b$from the \cite{DupreeRaymond83}.
\end{table*}

\begin{figure*}
\includegraphics[angle=90, width=\textwidth]{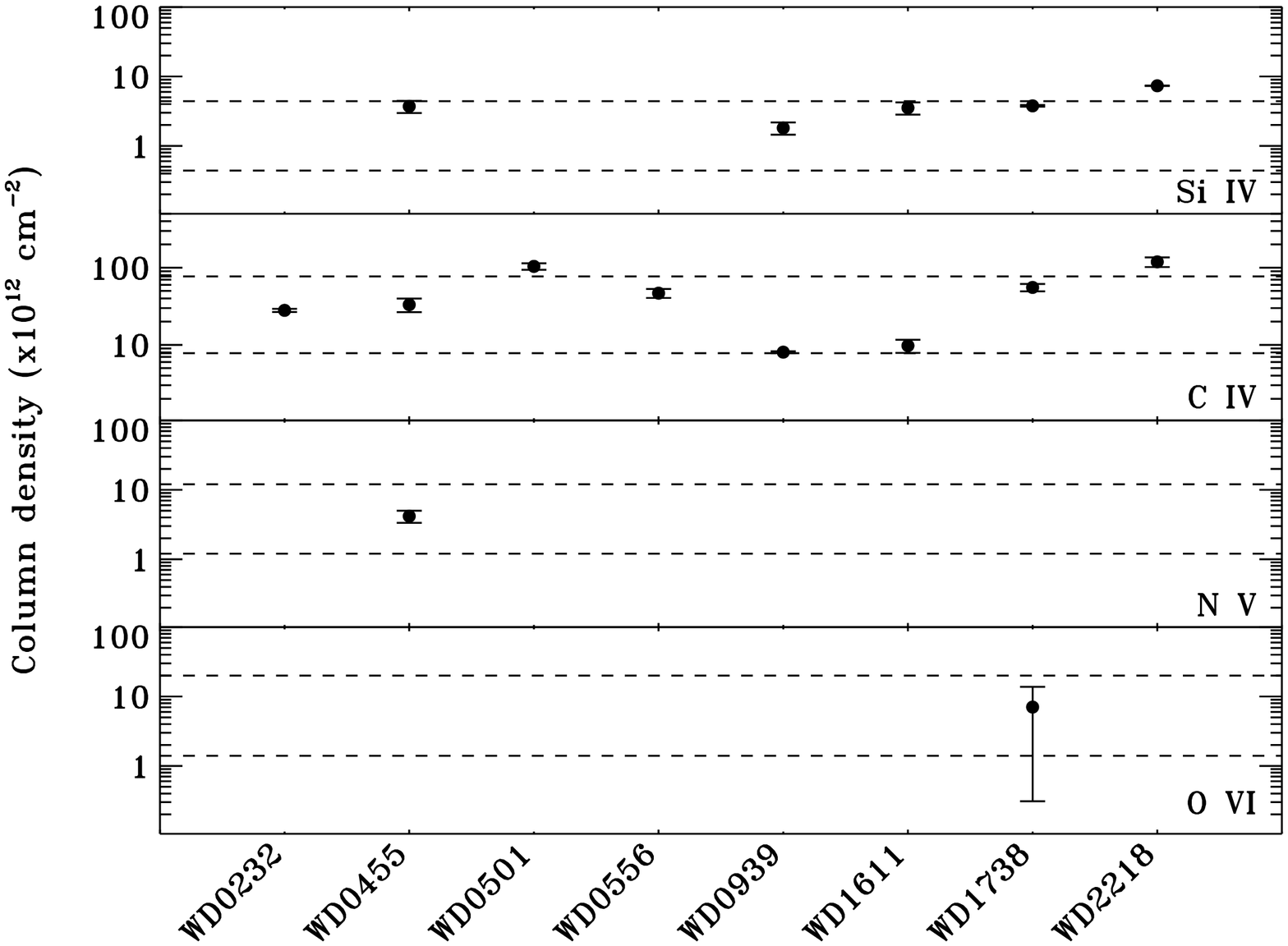}
  \caption{The column densities of circumstellar material reported in this study. The horizontal dashed lines are the boundaries of the predicted column density ranges for each ion by the \cite{DupreeRaymond83} DA Str\"{o}mgren sphere model.}
  \label{fig:col_strom}
\end{figure*}

The metal column density ranges given in table \ref{table:ISMcols} span a limited \textit{n}$_{\rm H}$ range; the full metal column density ranges predicted by this model are  0.44x10$^{12} <$ Si\,{\sc iv} $<$ 4.4x10$^{12} \rm cm^{-2}$, 7.8x10$^{12} <$ C\,{\sc iv} $<$ 77x10$^{12} \rm cm^{-2}$, 1.2x10$^{12} <$ N\,{\sc v} $<$ 12x10$^{12} \rm cm^{-2}$ and  1.4x10$^{12} <$ O\,{\sc vi} $<$ 20x10$^{12} \rm cm^{-2}$ (\citealt{DupreeRaymond83}) for the full range of \textit{n}$_{\rm H}$ values (table 4, \citealt{Bannister03}) along the sight lines to the stars in this study. On an object-to-object basis, the predicted column densities do not match those observed. However, the limits of the range of expected metal column densities are close to the observed range (figure \ref{fig:col_strom}). 

As discussed by \cite{Bannister03}, several limitations exist in the Str\"{o}mgren sphere models of \cite{DupreeRaymond83}. A single DA (and a DO) white dwarf was considered with \textit{T}$_{\rm eff}$\,=\,60\,000\,K and log\,\textit{g}\,=\,8.0. The \textit{T}$_{\rm eff}$ range for the white dwarfs with circumstellar material studied here varies from 38\,840\,K to 69\,711\,K; the increase in radiated flux above 60\,000\,K would increase the radius of the Str\"{o}mgren sphere (\textit{r}$_{\rm S}$), and thus ionise more ISM and increase the column density, while the white dwarfs with \textit{T}$_{\rm eff}$\,$<$\,60\,000\,K would have far smaller \textit{r}$_{\rm S}$ values. At the time of the \cite{DupreeRaymond83} study, a possible source of the high ion absorption lines seen in hot white dwarf spectra was thought to be the ionisation of circumstellar or ISM  material inside white dwarf Str\"{o}mgren spheres \citep[e.g.][]{BruhweilerKondo81}. However, over the past few decades of white dwarf research, it has been shown that these features mostly have a photospheric nature (e.g. \citealt{Barstowetal93}, \citealt{Chayeretal94}, \citealt{Chayeretal95}, \citealt{Chayeretal95b}, \citealt{Marshetal97}, \citealt{Barstowetal03}, Barstow et al. \textit{in preparation}, etc). The absorption lines due to photospheric metals act to reduce flux at specific wavelengths, lowering the radiated flux and therefore reducing  \textit{r}$_{\rm S}$; for \textit{T}$_{\rm eff}$\,$>$\,50\,000\,K the line blanketing due to Fe peak elements may make this effect more severe (though the flux redistributed by line blanketing may contribute to energies still above lower ionisation potentials important for the structure of the Str\"{o}mgren sphere). The ISM component in the model was assumed to be isothermal (at 40\,000\,K) and of uniform density. For all of these reasons, the model column densities in figure \ref{fig:col_strom} should be looked upon as a rough estimate of scale rather than precise predictions. 

The study of \cite{TatTerzian99} provided \textit{r}$_{\rm S}$ values for a selection of hot white dwarfs, including WD\,0232+035 and  WD\,0501+527, for \textit{n}$_{\rm e}$\,=\,0.01 cm$^{-3}$ and 0.03 cm$^{-3}$. \textit{r}$_{\rm S}$ estimates for the DAs with circumstellar features (table \ref{table:dist}) were made by adopting the \textit{r}$_{\rm S}$ values from a \cite{TatTerzian99} model star with a similar \textit{T}$_{\rm eff}$ to each of the stars with circumstellar absorption. Again, the \textit{r}$_{\rm S}$ values of \cite{TatTerzian99} were calculated using  the basic approach of \cite{Stromgren39}. A more thorough approach to estimating \textit{r}$_{\rm S}$ for the stars in this sample would be to model the interaction of the stellar flux distribution (taking proper account of photospheric metal absorption and the white dwarf UV spectral energy distribution) with the ISM (using measured ISM temperatures and electron and metal column densities), and is beyond the scope of this study.
 
Using the distances to the white dwarfs with circumstellar material and their \textit{r}$_{\rm S}$ estimates, a minimum distance to the material ionised by the white dwarf can be estimated (table \ref{table:dist}). Given that the  Str\"{o}mgren sphere of WD\,0232+035 comfortably contains its binary companion, it may also be the case that material inside the Str\"{o}mgren sphere, this time from mass lost from the companion, is being photoionised (this would explain the difference between $v$$_{\rm CS}$ and the ISM velocities at this object). 

\begin{table*}
\caption{The estimated \textit{r}$_{\rm S}$ (from \citealt{TatTerzian99}) for each of the white dwarfs in this sample that display circumstellar material, and the estimated minimum distances (\textit{D}) to the ISM component being ionised by the star (for \textit{n}$_{\rm e}$\,=\,0.01 and 0.03 cm$^{-3}$). The ISM component matching the circumstellar material is stated in column 6; a `?' signifies a tentative association to the Hyades cloud. The model star used to obtain the \textit{r}$_{\rm S}$ values are as detailed \citep[with the \textit{T}$_{\rm eff}$ assumed by ][]{TatTerzian99}. All distances are expressed in pc.}
\begin{tabular}{c c c c c c}
\hline
WD             &  \textit{r}$_{\rm S}$ (\textit{n}$_{\rm e}$\,=\,0.01 cm$^{-3}$) & \textit{D} (\textit{n}$_{\rm e}$\,=\,0.01 cm$^{-3}$) & \textit{r}$_{\rm S}$ (\textit{n}$_{\rm e}$\,=\,0.03 cm$^{-3}$) & \textit{D} (\textit{n}$_{\rm e}$\,=\,0.03 cm$^{-3}$) & ISM component\\
\hline
0232+035$^a$ & 33.66                                                           & 44.34                                                &  16.18                                                         & 61.82                                                        & \\
0455$-$282$^b$ & 26.00                                                           & 82                                                   &  12.50                                                         & 95.5                                                         & \\
0501+527$^b$ & 26.00                                                           & 24                                                   &  12.50                                                         & 37.5                                                        & ISM,pri (Hyades?)\\
0556$-$375$^a$ & 33.66                                                           & 261.34                                               &  16.18                                                         & 278.82                                                       & \\
0939+262$^c$ & 39                                                              & 178                                                  &  18.86                                                         & 198.14                                                       & \\
1611$-$084$^d$ & 18.43                                                           & 67.57                                                &  8.86                                                          & 77.14                                                        &\\
1738+669$^c$ & 39                                                              & 204                                                  &  18.86                                                         & 224.14                                                       & ISM,pri\\
2218+706$^a$ & 33.66                                                           & 402.34                                               &  16.18                                                         & 419.82                                                       & ISM,pri\\
\hline
\end{tabular}
\\Note: though \textit{r}$_{\rm S}$ estimates were made for WD\,0232+035 and WD\,0501+527, the \textit{T}$_{\rm eff}$ values assumed by \cite{TatTerzian99} are significantly different to those in table \ref{table:tloggobs}, so the \textit{r}$_{\rm S}$ value for a  model star with a \textit{T}$_{\rm eff}$ closer to those in table \ref{table:tloggobs} was used. An absence of value in column six signifies that the circumstellar material does not match any detected ISM component, so the distance is an estimate of the Str\"{o}mgren sphere boundary.  
\\$^a$WD\,0501+527 (\textit{T}$_{\rm eff}$\,=\,61\,160\,K), $^b$WD\,0232+035 (\textit{T}$_{\rm eff}$\,=\,50\,000\,K), $^c$WD 1211+332 (\textit{T}$_{\rm eff}$\,=\,70\,000\,K), $^d$WD\,2111+498 (\textit{T}$_{\rm eff}$\,=\,39\,800\,K). 
\label{table:dist}
\end{table*}

The examination of the relationship between circumstellar line shift and \textit{T}$_{\rm eff}$ conducted by \cite{Lallementetal11} for their sample and the samples of \cite{HBS98}, \cite{Bannister03} and \cite{Barstowetal10}, found considerable spread in the line shifts around a trend of decreasing line shift with \textit{T}$_{\rm eff}$, attributed to the inclusion of different types of white dwarf/circumstellar environments. This highlights another source for the possible dispersion in velocity shift seen by \cite{Lallementetal11}; the variety of $v$$_{\rm phot}$ values and ISM cloudlet velocities will give rise to a variety of line shift values. 

Previous studies that looked at the distribution of high ions in the ISM may have come to incorrect conclusions about the structure of hot interstellar gas, if the non-photospheric high ions observed here are due to the ionisation of the ISM by hot white dwarfs. \cite{Barstowetal10} and \cite{SL06} cite ISM interfaces as sources of non-photospheric O\,{\sc vi}; in the cases where the non-photospheric high ions are due to the ionisation of ISM by hot white dwarfs, the O\,{\sc vi} can be associated with the Str\"{o}mgren spheres of the stars rather than such interfaces. Indeed, \cite{Lallementetal11} found that conductive interfaces cannot explain the observed circumstellar material seen in their sample, since the line widths of C\,{\sc iv} and O\,{\sc vi} are too small, the C\,{\sc iv} and O\,{\sc vi} columns are not what are expected and no predicted dynamical link between the observed C\,{\sc iv} and O\,{\sc vi} is seen. \cite{Welshetal10} found C\,{\sc iv} within the local cavity, with a Doppler width narrower than that predicted by interface regions, but consistent with photoionisation by B stars, demonstrating the viability of hot stars ionising ISM in their locality to explain the observed high ions. This shows that previous studies may not have correctly interpreted the source of non-photospheric high ions in DA spectra, and thus the current view of the distribution of hot interstellar gas may not be wholly correct, for example the lack of correlation between O\,{\sc vi} and the soft X-ray background (SXRB) found by  \cite{Barstowetal10} shows a more localised O\,{\sc vi} source may be present, rather than a global ISM phenomenon.  The conclusion of \cite{Welshetal10b} that the structure of the ionised ISM is cellular, due to photoionisation by B stars and hot white dwarfs, lends weight to this argument.

 This shows clearly the importance of the white dwarf circumstellar environment to our understanding of the ISM, and demonstrates the need for a detailed analysis of each star in which circumstellar high ion features are seen, to ascertain the source of the material which is photoionised by the white dwarf, and thus better inform our understanding of the physical structure of the ISM and the photoionisation of it by hot white dwarf stars.

\subsection{Mass loss}
\label{Mass loss}
\begin{figure}
\includegraphics[angle=90, width=0.47\textwidth]{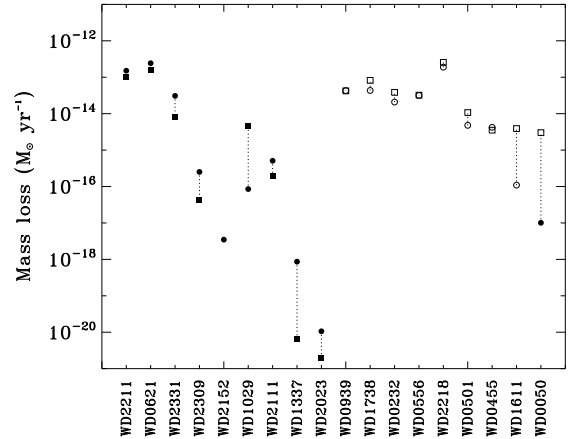}
  \caption{A comparison of the mass loss rates calculated by \cite{Bannister03} (squares) to those computed with updated metal abundances (circles). The stars without circumstellar material are plotted with filled symbols, while those with circumstellar material are plotted with open symbols. The stars are ordered by \textit{T}$_{\rm eff}$, with the hottest stars to the left and the cooler stars to the right, to allow the trend found by \cite{Bannister03} to be observed. The data points for each star are joined with a dotted line to aid comparison.}
  \label{fig:mdot_nb}
\end{figure}

Previous work, e.g. \cite{HBS98}, attributed the observed circumstellar high ions to stellar mass loss. Indeed, using the \cite{Abbott82} mass loss formalism, \cite{Bannister03} reported that all stars with circumstellar material had a high mass loss rate, whereas the stars without circumstellar material had a decreasing mass loss rate with decreasing \textit{T}$_{\rm eff}$ (figure \ref{fig:mdot_nb}, square symbols). Using up to date white dwarf metal abundances from \cite{Dickinson11} and Barstow et al. (\textit{in preparation}), and solar abundances from \cite{Asplundetal09},  the previous mass loss pattern is no longer seen (figure \ref{fig:mdot_nb}, circular symbols). Indeed, the star with the highest mass loss rate calculated using the \cite{Abbott82} formalism (WD\,0621$-$376) does not have any non-photospheric high ion absorption.

The formula of \cite{Abbott82} is based on the parameterisation of the dependence of the radiative force on the wind optical depth parameter, which is valid in a limited range only. This parameterised form predicts an increasing radiative force with decreasing wind density. However, it does not take into account that at some point (depending on metallicity) the radiative force will saturate. Therefore, the formula always provides a non-zero mass loss rate, even for such compact stars in which the radiative force cannot exceed the gravitational force.

In more recent studies of mass loss in hot white dwarf stars, it has been found that for DAs with 25\,000\,K\,$<$\,\textit{T}$_{\rm eff}$\,$<$\,50\,000\,K no winds can exist for log\,\textit{g}\,$>$\,7.0 for solar or sub-solar metal abundances, since the radiative acceleration saturates below the gravitational acceleration of the star; likewise for a DA with \textit{T}$_{\rm eff}$\,=\,50\,000\,K and metal abundances 10$^{-2}$ times the solar value, no winds can exist above a log\,\textit{g} of 4.5 \citep{Unglaub08}. For a white dwarf with \textit{T}$_{\rm eff}$\,$=$\,60\,000\,K, it has also been found that mass loss cannot occur for log\,\textit{g}\,$>$\,7.0 \citep{Unglaub07}. Given that the stars studied here have metal abundances 10$^{-2}$-10$^{-4}$ times the solar abundances, and that the lowest log\,\textit{g} of the stars studied here is 7.05, it can reasonably be concluded that mass loss cannot account for the observed circumstellar material seen in the spectra of WD\,0455$-$282, WD\,0501+527, WD\,0556$-$375, WD\,1611$-$084 and WD\,2218+706. The remaining white dwarfs with circumstellar detections (WD\,0232+035, WD\,0939+262 and WD\,1738+669) are not much hotter than 60\,000\,K. Though no mass loss calculations have been made for \textit{T}$_{\rm eff}$\,$>$\,60\,000\,K, the existence of winds also seems to be unlikely in hotter DAs with subsolar metal abundances. In thin winds, the major contribution to the radiative force comes from the strong lines of the CNO elements \citep{Vinketal01}. When, however, more and more particles of these elements change into the helium-like stage of ionisation at high \textit{T}$_{\rm eff}$, the radiative force is reduced, because the lines of this ionisation stage are at very short wavelengths outside the flux maximum. So, in spite of their higher luminosity, it is not clear that winds are more likely to exist at DAs with \textit{T}$_{\rm eff}$\,$>$\,60\,000\,K than at cooler ones.

If winds exist in hot DA white dwarfs at all, they should consist of metals only, since hydrogen is in hydrostatic equilibrium. Initial calculations for a DA with \textit{T}$_{\rm eff}$\,=\,66\,000\,K with log\,\textit{g}\,=\,7.7 show that the outward flow of the elements C, N and O in an otherwise hydrostatic stellar atmosphere cannot exceed a value of the order 10$^{-19}$ M$_{\odot}$ yr$^{-1}$. This maximum value can be derived due to the dependence of the radiative force on the abundance of the element, using arguments similar to those of \cite{Seaton96}. It indirectly follows that the mass loss rate of these elements cannot be significantly higher than 10$^{-19}$ M$_{\odot}$ yr$^{-1}$, because the stellar atmosphere would otherwise be rapidly emptied. As shown by \cite{Votrubaetal10} for the case of sdB stars, metallic winds are accelerated to velocities of several thousands of\,km\,s$^{-1}$, exceeding the escape velocity of the star. This suggests that even if mass loss is occurring, the  metals would form a thin wind, and would not be able to explain the circumstellar material seen here.

\subsection{Ancient planetary nebulae}
\label{Ancient planetary nebulae}

\begin{figure}
\includegraphics[angle=90, width=0.47\textwidth]{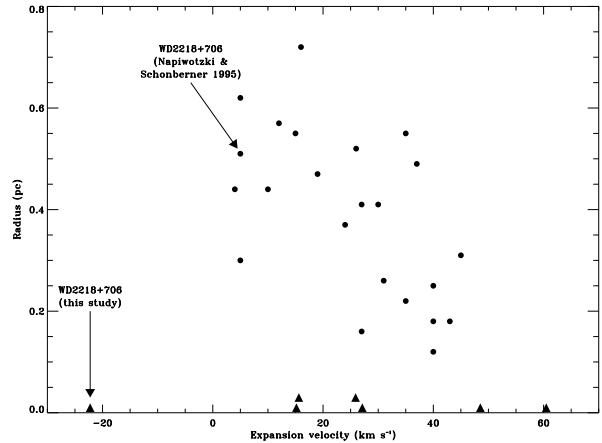}
  \caption{The planetary nebula expansion velocities ($v$$_{\rm exp}$) from the \cite{NapiwotzkiSchonberner95} sample (circles) plotted with the $v_{\rm phot} - v_{\rm circ}$ values of the stars in this study (triangles). Since there are no nebula radius measurements for the stars here, they are plotted at 0\,km\,s$^{\rm -1}$. Overlapping $v_{\rm phot} - v_{\rm circ}$ values are offset for clarity.}
  \label{fig:vexp_fin}
\end{figure}

Given that one of the stars (WD\,2218+706) is a bona fide CSPN, it is sensible to examine whether the observed circumstellar material can be associated with ancient, diffuse PNe. Following the approach of \cite{Bannister03}, the PN expansion velocities ($v$$_{\rm exp}$) from \cite{NapiwotzkiSchonberner95} were compared to the $v_{\rm phot} - v_{\rm circ}$ values (figure \ref{fig:vexp_fin}). Given the broad consistency between the velocities measured here and those measured by \cite{Bannister03}, it is perhaps not surprising this comparison yields roughly the same result.  The range of $v$$_{\rm exp}$ values is broadly matched by the range of $v_{\rm phot} - v_{\rm circ}$ values, with a few outliers. Again, the $v_{\rm phot} - v_{\rm circ}$ of WD\'2218+706 is far from its $v$$_{\rm exp}$, indicating that the non-photospheric high ions seen here may no longer be associated with the PN. Though \cite{Bannister03} also used the presence of the PN TK2 at WD\,1738+669 to explain the circumstellar high ions seen at this star, \cite{FrewParker06} found that this object is not actually a CSPN, citing the ionisation of material in the Str\"{o}mgren sphere of the white dwarf as giving the impression of a nearby PN. Coupled with the difference in $v$$_{\rm exp}$ and  $v_{\rm phot} - v_{\rm circ}$ at WD\,2218+706 and the similarity in $v$$_{\rm CS}$ and $v$$_{\rm ISM, pri}$ at both WD\,1738+669 and WD\,2218+706, it seems that the circumstellar material seen at WD\,2218+706 may, again, in fact be ionised ISM in the Str\"{o}mgren sphere of the object (though this material may have originated in the PN at the star). 

\subsection{Summary}
\label{Summary}

We have re-examined the hot white dwarf sample of \cite{Bannister03} using an improved measurement technique, which for the first time has provided column densities for all of the observed circumstellar material. Unambiguous  circumstellar detections are made at WD\,0232+035, WD\,0455$-$282, WD\,0501+527, WD\,0556$-$375, WD\,0939+262, WD\,1611$-$084, WD\,1738+669 and WD\,2218+706. Several sources for this circumstellar material (circumstellar discs, ionised ISM, stellar mass loss and ancient PNe) were examined.

Once thought to be a significant potential mechanism for the production hot DA circumstellar features, mass loss was found not to be a plausible explanation for the observations here. Similarly, ancient PNe at the stars are unlikely to be the source of the circumstellar absorption seen in the spectra of WD\,1738+669 and WD\,2218+706, as was once thought. 

The `circumstellar' absorption seen in the spectrum of WD\,0501+527 has previously been attributed to the ionisation of the Hyades cloud by the hot white dwarf; this conclusion remains intact in this work. While the ionisation of the ISM near WD\,1738+669 and WD\,2218+706 may be a credible source of the observed circumstellar features, some difficulty is met in linking the ionisation of the ISM to the circumstellar features at WD\,0455$-$282, WD\,0556$-$375 and WD\,0939+262. At these stars, $v$$_{\rm CS}$ does not match up well with the detected ISM components. Some agreement is seen between $v$$_{\rm CS}$ and $v$$_{\rm LISM,pred}$ along the sight lines to these stars, though the stars are too far away to ionise the LISM. It may therefore be the case that ISM cloudlets near the stars with velocities similar to those in the LISM are in fact being ionised. The measured column densities of the circumstellar material also lie within the range of ionised ISM column densities predicted by the \cite{DupreeRaymond83} DA Str\"{o}mgren sphere model.

It could also be, as suggested by \cite{Lallementetal11}, that these stars are ionising either the evaporated remains of circumstellar rocky bodies or analogues to the circumstellar discs seen at cooler white dwarf stars. Indeed, the $v$$_{\rm CS}$ of WD\,1614$-$084 does not match any interstellar component, predicted or detected. Though the $v$$_{\rm CS}$ at WD\,0232+035 does not line up with any measured or predicted ISM components, the Str\"{o}mgren sphere of the object is more than large enough to ionise any material lost from its dwarf companion.

The potential implications of this to our understanding of the ISM are significant. Where previous studies attributed non-photospheric high ions to hot/cold gas interfaces, a picture of the distribution of the ISM has built up. Should at least some of the non-photospheric high ions observed have arisen in hot white dwarf Str\"{o}mgren spheres, the conclusions drawn as to the structure of the ISM using hot/cold gas interface models would not be wholly correct, and our understanding of the ionised ISM will need to be revised. Similarly, should at least some of the circumstellar high ions here be due to the ionisation of circumstellar material from evaporated planetesimals/disrupted extrasolar minor planets, studies of systems such as these will provide a valuable insight into the evolution of extrasolar planetary systems through the hot white dwarf phase.

This work clearly shows the importance of understanding the white dwarf circumstellar environment to our understanding of the ISM and the end states of stellar and planetary system evolution, and provides a good case for the re-observation of stars where unresolved circumstellar material may be resident with higher resolution instruments to identify new cases of circumstellar absorption. High resolution re-observation of the stars with circumstellar material here will also allow a better characterisation of the ISM along the sight lines to the stars. A detailed model of hot white dwarf Str\"{o}mgren spheres, including advances in both our knowledge of hot white dwarf atmospheres and the ISM, is clearly required to better understand this phenomenon. Physical modelling of the interaction of planetesimals and circumstellar discs with the hot stars is a crucial future step to better understanding the evolution of extrasolar planetary systems at the hot white dwarf stage, and may provide testable predictions that can be compared to the quantities observed here.

%%%%%%%%%%%%%%%%%%%%%%%%%%%%%%%%%%%%%%%%%%%%%%%%%%%%%%%%%%%%%%%%%%%%%%%%%%%%%%
%% Appendix. Comments on individual stars.                                  %%
%%%%%%%%%%%%%%%%%%%%%%%%%%%%%%%%%%%%%%%%%%%%%%%%%%%%%%%%%%%%%%%%%%%%%%%%%%%%%%
\section{Comments on individual stars.}
\label{appendix}

Details of the measurements made for each star are discussed here. Representative plots are shown in the interesting case of WD\,0948+534 (section \ref{PG0948+534}).

\subsection{WD\,0050$-$335 (GD\, 659)}
\label{GD659}

\cite{Bannister03} found that a circumstellar component may be present in the coadded C\,{\sc iv} doublet at  $-$2.97$\pm$3.00\,km\,s$^{-1}$, while $v$$_{\rm phot}$\,=\,34.28$\pm$0.27\,km\,s$^{-1}$. The secondary component was relatively weak (6 m\AA) when compared to the photospheric component (36 m\AA). Though an \textit{F}-test showed this second component was statistically preferred, it was similar to nearby noise features. Given the similarity of this previously identified circumstellar feature to the noise, and the inability to coadd doublet components in velocity space here, a circumstellar measurement is not made. The detected low ion ISM component along the line of sight to this star is close to the expected velocity of the LIC, and may be associated with it.

\subsection{WD\,0232+035 (Feige\,24)}
\label{Feige24}

The data were obtained at two phases of the binary cycle, 0.73-0.74 (29$^{\rm th}$ November 1997) and 0.23-0.25 (4$^{\rm th}$ January 1998). The photospheric velocity varies between data sets, with $v$$_{\rm phot}$\,=\,30.11$\pm$0.52\,km\,s$^{-1}$ and 128.23$\pm$0.31\,km\,s$^{-1}$ at each binary phase, respectively. In the C\,{\sc iv} doublet, a second, stationary set of absorption features is seen at $v$$_{\rm CS}$\,=\,7.4$\pm$0.34\,km\,s$^{-1}$. The column density of this component is (2.8 $\pm$ 0.13)x10$^{13}$\,cm$^{-2}$, and it has a \textit{b} value of 6.4$\pm$0.5\,km\,s$^{-1}$. A single, photospheric component in the O\,{\sc vi} absorption lines is seen, consistent with the findings of \cite{Barstowetal10}.

The Si\,{\sc ii} and S\,{\sc ii} lines give a $v$$_{\rm ISM, pri}$\,=\,2.85$\pm$0.34\,km\,s$^{-1}$ and $v$$_{\rm ISM, sec}$\,=\,17.1$\pm$1.3\,km\,s$^{-1}$. The projected $v$$_{\rm LISM}$ due to the LIC is found to be 18.1$\pm$1.13\,km\,s$^{-1}$, near $v$$_{\rm ISM, sec}$ but far from both $v$$_{\rm ISM, pri}$ and $v$$_{\rm CS}$.  Given the white dwarf has a M dwarf companion, it is possible that the observed circumstellar features are due to the ionisation of the mass lost from the binary companion. Indeed \cite{Kawkaetal08} reasoned that the observed O\,{\sc vi} abundances seen in their post common envelope binary sample (which included WD\,0232+035) are due to the accretion of mass lost from binary companions, given the O\,{\sc vi} reservoir at the top of the photospheres of the white dwarf sample of \cite{Chayeretal06}. It was, however, stated that the relatively large orbital separation of WD\,0232+035 \citep[\textit{p}\,=\,4.23\,d, ][]{VennesThorstensen94} makes accretion from the binary companion much weaker than for the other binaries in their sample (with 0.33 $<$ \textit{p} $<$ 1.26\,d). 

\subsection{WD\,0455$-$282 (REJ\,0457$-$281)}
\label{REJ0457$-$281}

Blueshifted features in the UV spectrum of this object were first reported in the C\,{\sc iv} and Si\,{\sc iv} absorption line profiles \citep{HBS98}. Some evidence for circumstellar material in the 1239\,\AA\ N\,{\sc v} absorption line was noted by \cite{Bannister03}. Here, $v$$_{\rm phot}$\,=\,79.28$\pm$1.79\,km\,s$^{-1}$, and the C\,{\sc iv}, N\,{\sc v} and Si\,{\sc iv} circumstellar components are found at an average velocity of 18.8$\pm$3.47\,km\,s$^{-1}$, with column densities of (3.32$\pm$0.66)x10$^{13}$\,cm$^{-2}$, (4.17$\pm$0.84)x10$^{11}$\,cm$^{-2}$ and (3.71$\pm$0.74)x10$^{12}$\,cm$^{-2}$, respectively. The line of sight to the star traverses the Blue cloud (which has a projected velocity of 12.56$\pm$1.03\,km\,s$^{-1}$), though this star is too far away from the cloud to account for the observed circumstellar absorption. Circumstellar material or unresolved ISM may account for this observation. 

\subsection{WD\,0501+527 (G191-B2B)}
\label{G191$-$B2B}

WD\,0501+527 is one of the best studied hot white dwarfs. In keeping with the results of \cite{Bannister03}, circumstellar C\,{\sc iv} is seen at 8.9$\pm$0.07\,km\,s$^{-1}$ and the  averaged $v$$_{\rm phot}$\,=\,24.51$\pm$0.16\,km\,s$^{-1}$. \textit{N}(C\,{\sc iv})\,=\,(1.04$\pm$0.1)x10$^{14}$\,cm$^{-2}$, and a \textit{b} value of 5.65$\pm$0.18\,km\,s$^{-1}$ \citep[near that of ][]{VennesLanz01} is measured for the circumstellar material. Two components are measured in the ISM, with the primary at 8.5$\pm$0.18\,km\,s$^{-1}$ and the secondary at 19.3$\pm$0.03\,km\,s$^{-1}$, agreeing with the values found by \cite{Sahuetal99} and \cite{RedfieldLinsky04}. Two LISM clouds are traversed by the line of sight to WD\,0501+527. The Hyades cloud has a projected velocity of 9.35$\pm$1.32\,km\,s$^{-1}$, matching both $v$$_{\rm CS}$ and $v$$_{\rm ISM, pri}$, suggesting that the primary ISM component and the observed circumstellar material may reside in this cloud; the electron density of the Hyades cloud along the line of sight to WD\,0501+527 supports this association (\citealt{RedfieldFalcon08}). However, though  \cite{RedfieldLinsky04} also suggest the absorption seen at $v$$_{\rm CS}$ is due to the photoionisation of the Hyades cloud by WD\,0501+527, the metal depletion value of this line of sight is anomalous when compared to the other sight lines through the cloud, signalling some caution must be observed here. It may be the case that some material with a velocity similar to the Hyades cloud, closer to the star (at a distance of 50\,pc) is in fact being ionised. The LIC, with  a projected velocity of 19.1$\pm$1.1\,km\,s$^{-1}$, is likely to be responsible for the observed secondary ISM component. The O\,{\sc vi} 1032\,\AA\ line has a single component at 19$\pm$2.3\,km\,s$^{-1}$ and is attributed to the photosphere, in agreement with both \cite{SL06} and \cite{Barstowetal10}.

\subsection{WD\,0556$-$375 (REJ\,0558$-$373)}
\label{REJ0558$-$373}

$v$$_{\rm phot}$\,=\,25.37$\pm$2.03\,km\,s$^{-1}$. Circumstellar material is again seen in the C\,{\sc iv} doublet (giving $v$$_{\rm CS}$\,=\,10.2$\pm$1.07\,km\,s$^{-1}$), with an averaged column density of (4.67$\pm$0.62)x10$^{13}$\,cm$^{-2}$ and \textit{b}\,=\,11$\pm$1.3\,km\,s$^{-1}$. Three ISM components are seen in the ISM detected at 7.8$\pm$1\,km\,s$^{-1}$, 19.9$\pm$1.7\,km\,s$^{-1}$ and 36$\pm$2.3\,km\,s$^{-1}$ (ordered by equivalent width). The sight line to WD\,0556$-$375 traverses the Blue cloud, which has a projected velocity of 11.36$\pm$0.95\,km\,s$^{-1}$, over-lapping with $v$$_{\rm CS}$ within errors, and near $v$$_{\rm ISM, pri}$. However, like the other stars with circumstellar absorption, this star lies at too great a distance (295\,pc) to ionise the LISM; either another ISM cloud with a similar velocity to the Blue cloud is present near enough to the star to facilitate photoionisation, or circumstellar material of the type describe by \cite{Lallementetal11} is present.

\subsection{WD\,0621$-$376 (REJ\,0623$-$371)}
\label{REJ0623$-$371}

No clear evidence for circumstellar material is seen at this star. $v$$_{\rm phot}$ is measured at 39.44$\pm$0.25\,km\,s$^{-1}$, while a single component ISM fit gives $v$$_{\rm ISM}$\,=\,15.8$\pm$0.4\,km\,s$^{-1}$. As with WD\,2211$-$495, the inconsistency seen in the centroid positions of the C\,{\sc iv} doublet components ($v$$_{1548}$\,=\,37.6$\pm$0.6\,km\,s$^{-1}$, $v$$_{1550}$\,=\,48.5$\pm$0.7\,km\,s$^{-1}$) for this star was used to indicate the possible presence of unresolved circumstellar material by \cite{Bannister03}. Indeed, the relatively poor resolution of the \textit{IUE}[SWP] data may hide the shifted circumstellar material, and higher resolution data may allow this material to be resolved.

\subsection{WD\,0939+262 (Ton\,021)}
\label{Ton021}

The C\,{\sc iv} and Si\,{\sc iv} doublets both display circumstellar material, giving $v$$_{\rm CS}$\,=\,9.38$\pm$6.59\,km\,s$^{-1}$. When the N\,{\sc v} and O\,{\sc v} lines are included, $v$$_{\rm phot}$\,=\,36.5$\pm$0.47\,km\,s$^{-1}$. The average column density for the C\,{\sc iv} doublet is (8.05$\pm$0.21)x10$^{12}$\,cm$^{-2}$ with a \textit{b} value of 8.3$\pm$1.75\,km\,s$^{-1}$; for Si\,{\sc iv} it is (1.81$\pm$0.36)x10$^{12}$\,cm$^{-2}$ with a \textit{b} value of 11.65$\pm$5.28\,km\,s$^{-1}$. The ISM is found to have one component at $-$2.1$\pm$0.2\,km\,s$^{-1}$. Though $v$$_{\rm LISM,pred}$ (due to the LIC) is predicted to be at 10.81$\pm$1.29\,km\,s$^{-1}$, (lining up well with $v$$_{\rm CS}$) for reasons discussed for WD\,0455$-$282 and WD\,0556$-$375 this LISM cloud cannot be ionised by this star, implying that the observed circumstellar material may reside in a `LIC-like' cloud or in a circumstellar disc.

\subsection{WD\,0948+534 (PG\,0948+534)}
\label{PG0948+534}

\begin{figure}
\includegraphics[width=0.47\textwidth]{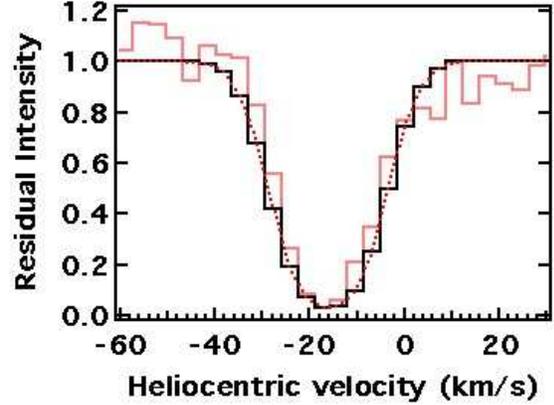}
  \caption{The 1548\,\AA\ C\,{\sc iv} line of WD\,0948+534, fit with a single absorbing component at $-$16\,km\,s$^{-1}$. The dark solid line shows the model, while the dotted line signifies the model components. The lighter solid line is the observed data. This plotting convention is used in all subsequent line profile plots.}
  \label{fig:pg0948_CIV_single}
\end{figure}

Three ISM components are found here, in keeping with the results of \cite{Bannister03}. The primary component is at $-$18.45$\pm$0.42\,km\,s$^{-1}$, with the secondary and tertiary components at  $-$1.60$\pm$0.63\,km\,s$^{-1}$ and 22.6$\pm$0.8\,km\,s$^{-1}$. In contrast, \cite{Bannister03}  found the $-$1.6\,km\,s$^{-1}$ component had an equivalent width larger than that of the $-$18.45\,km\,s$^{-1}$ component.  In the S\,{\sc ii} line, the model equivalent widths are similar at 32.14 m\AA\ and 30.99 m\AA\ for the $-$18.45\,km\,s$^{-1}$ and $-$1.6\,km\,s$^{-1}$ components, respectively. However, the Si\,{\sc ii} line has model equivalent widths of 39.20 m\AA\ and 25.63 m\AA\ for the $-$18.45\,km\,s$^{-1}$ and $-$1.6\,km\,s$^{-1}$ components, clearly making the $-$18.45\,km\,s$^{-1}$ component the primary. One absorbing component is statistically preferred for the high ions (figure \ref{fig:pg0948_CIV_single}), with an averaged $v$$_{\rm phot}$ of  $-$17.09$\pm$1.73\,km\,s$^{-1}$. Caution must be exercised here; the modelling technique simply fit the absorption features with Gaussian profiles, and does not include physically robust stellar absorption line profiles. Indeed, modelling this object with a single component stellar model has proved difficult \citep{Dickinson11}.

\begin{figure}
\includegraphics[width=0.47\textwidth]{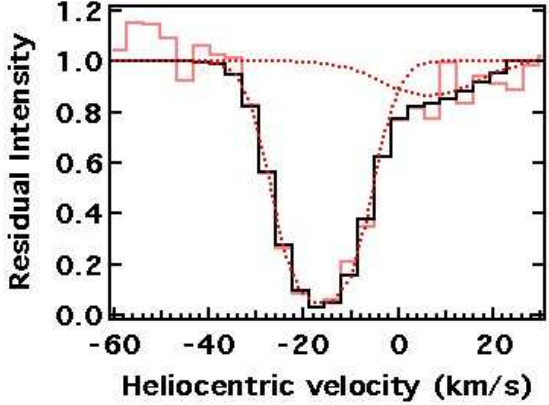}
  \caption{The 1548\,\AA\ C\,{\sc iv} line of WD\,0948+534, fit with two absorbing components at $v$$_{\rm phot, CIV}$\,=\,$-$17.6\,km\,s$^{-1}$ and $v$$_{\rm circ, CIV}$\,=\,1.65\,km\,s$^{-1}$.}
  \label{fig:pg0948_CIV_double}
\end{figure}

\begin{figure}
\includegraphics[width=0.47\textwidth]{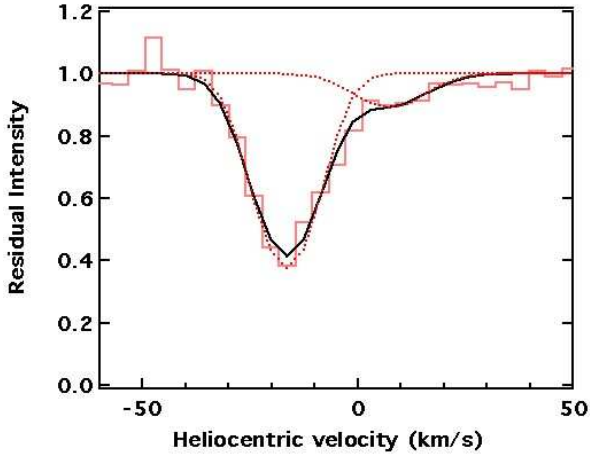}
  \caption{The 1393\,\AA\ Si\,{\sc iv} line of WD\,0948+534, fit with two absorbing components at $v$$_{\rm phot,SiIV}$\,=\,$-$16.9\,km\,s$^{-1}$ and $v$$_{\rm circ,SiIV}$\,=\,3.5\,km\,s$^{-1}$.}
  \label{fig:pg0948_SiIV_double}
\end{figure}

Considerable spread is seen between the high ions ($v$$_{\rm OV}$\,=\,$-$10.03\,km\,s$^{-1}$, $v$$_{\rm CIV}$\,=\,$-$16\,km\,s$^{-1}$). \cite{Bannister03} used the inconsistencies in the centroid positions of the absorption features in WD\,0621$-$376 and WD\,2211$-$495 to infer the possible existence of unresolved circumstellar material in the spectra of the stars. Though not statistically preferred, a hint of  a secondary component can be seen in the C\,{\sc iv} doublet (most obviously in the 1548\,\AA\ line). Fitting with two components gives $v$$_{\rm phot, CIV}$\,=\,$-$17.6$\pm$0.39\,km\,s$^{-1}$ and $v$$_{\rm circ, CIV}$\,=\,1.65$\pm$7.22\,km\,s$^{-1}$ (figure \ref{fig:pg0948_CIV_double}). The large error on $v$$_{\rm circ, CIV}$ can be put down to the fact that the circumstellar material is not resolved. Applying this to the N\,{\sc v}, O\,{\sc v} and Si\,{\sc iv} lines (figure \ref{fig:pg0948_SiIV_double}) gives an averaged $v$$_{\rm phot}$ of $-$17.09$\pm$1.72\,km\,s$^{-1}$ and $v$$_{\rm CS}$\,=\,0.2$\pm$5.40\,km\,s$^{-1}$. An examination of the 1031.912 O\,{\sc vi} line in the \textit{FUSE} data also reveals two possible components at $-$22.9$\pm$1.2\,km\,s$^{-1}$ and 15.7$\pm$2.8\,km\,s$^{-1}$. Given that the \textit{FUSE} velocity resolution ($\sim$15\,km\,s$^{-1}$) is poorer than that of the \textit{STIS} E140M data ($\sim$7\,km\,s$^{-1}$) and that the absolute velocity calibration may not be the same for both the \textit{STIS} and \textit{FUSE} data, the secondary component at 15.7\,km\,s$^{-1}$ is deemed the circumstellar component. Using the \textit{STIS} measurements, $v$$_{\rm CSshift}$\,=\,$-$17.29$\pm$5.7\,km\,s$^{-1}$ and $v_{\rm ISM, sec}-v_{\rm phot}$\,=\,$-$18.69$\pm$1.8\,km\,s$^{-1}$, implying that if it is present, the circumstellar material may be related to the secondary ISM component. 

However tempting the presence of circumstellar material may be in explaining both the difficulty in modelling the absorption line profiles of the star and the inconsistency in the centroid positions of the high ions, one absorbing component is still statistically preferred. Higher resolution data may shed further light on this enigmatic object.

\subsection{WD\,1029+537 (REJ\,1032+532)}
\label{REJ1032+532}

In good agreement with the $v$$_{\rm phot}$\,=\,38.16$\pm$0.40\,km\,s$^{-1}$ found by \cite{Bannister03}, a value of  $v$$_{\rm phot}$\,=\,37.98$\pm$0.21\,km\,s$^{-1}$ is found here. $v$$_{\rm ISM}$\,=\,0.95$\pm$0.79\,km\,s$^{-1}$, again matching up with the value of 0.84$\pm$0.21\,km\,s$^{-1}$ found by \cite{Bannister03}. No circumstellar material is seen at this star.

\subsection{WD\,1057+719 (PG\,1057+719)}
\label{PG1057+719}

No evidence for circumstellar material is found in the spectrum of this white dwarf, nor is there any photospheric C\,{\sc iv}, N\,{\sc v}, O\,{\sc v} or Si\,{\sc iv}. A value of $v$$_{\rm ISM}$\,=\,$-$0.2$\pm$1.0\,km\,s$^{-1}$ is obtained.

\subsection{WD\,1123+189 (PG\,1123+189)}
\label{PG1123+189}

Using the S\,{\sc ii} and O\,{\sc i} lines, ISM components are seen at $-$4.75$\pm$3.18\,km\,s$^{-1}$ and 2.15$\pm$2.96\,km\,s$^{-1}$. The high uncertainties on these measurements are due to the saturation of the O\,{\sc i} line. No photospheric absorption features are seen. The C\,{\sc iv}, O\,{\sc v} and Si\,{\sc iv} lines are not covered by the \textit{STIS}[E140H] data, which covers 1160\,\AA\ - 1360\,\AA. Again, the N\,{\sc v} doublet cannot be coadded to obtain a $v$$_{\rm phot}$ estimate in the same way as it was by \cite{Bannister03}. 

\subsection{WD\,1254+223 (GD\,153)}
\label{GD153}

WD\,1254+223 is another white dwarf in which no C\,{\sc iv}, N\,{\sc v}, O\,{\sc v} and Si\,{\sc iv} absorption features are seen. A $v$$_{\rm ISM}$ measurement of $-$15.4$\pm$1.6\,km\,s$^{-1}$ is obtained, consistent with the values of both \cite{Bannister03} and \cite{HBS98}; these values disagree with the $v$$_{\rm ISM}$\, values around $-$5\,km\,s$^{-1}$ found by \cite{RedfieldLinsky04}, probably due to the use of higher resolution data in the latter study. O\,{\sc vi} was seen in the \textit{FUSE} spectrum  and is attributed to the ISM, in keeping with the findings of \cite{Barstowetal10}.

\subsection{WD\,1314+493 (HZ\,43)}
\label{HZ43}

Neither photospheric nor circumstellar absorption lines are seen in the spectrum of this DA. The $v$$_{\rm ISM}$ measured here using the low ionisation lines ($-$6.6$\pm$0.1\,km\,s$^{-1}$) is similar to the value found by \cite{RedfieldLinsky02} and is close to the projected velocity ($-$6.15$\pm$0.74\,km\,s$^{-1}$) of the NGP cloud along the line of sight to the star.

\subsection{WD\,1337+705 (EG\,102)}
\label{EG102}

With a \textit{T}$_{\rm eff}$\,=\,22\,090\,K, it is perhaps not surprising that no photospheric high ions are observed in the spectrum of WD\,1337+705. Though no circumstellar material is detected, WD\,1337+705 is another white dwarf in which the accretion of circumstellar material can  be used to explain the observed photospheric metal abundances. \cite{HolbergBarstowGreen97} found Mg\,{\sc ii} and Si\,{\sc ii} in the optical spectrum of WD\,1337+705, with Al\,{\sc ii} and Al\,{\sc iii} later detected in the \textit{IUE} spectrum \citep{HBS98}. The Al abundance (1.5x10$^{-8}$) measured by \cite{Bannister03} is in excess of that predicted by \cite{Chayeretal95}. \cite{HBS98} also found a C\,{\sc ii} line with a velocity that compared poorly both to the photosphere and the ISM along the sight line of the star.  While recent radiative levitation studies of objects in this temperature range predict some C, Al and Si in the photosphere of WD\,1337+705, accretion must still be occurring to explain the observed photospheric abundances (\citealt{ChayerDupuis10}, \citealt{Dupuisetal10}). \cite{ZuckermanReid98} detected a significant amount of Ca in the optical spectrum of WD\,1337+705 (Ca/H\,=\,2.5x10$^{-7}$), often used to infer the accretion of terrestrial like extrasolar planet remnants (e.g. \citealt{Zuckermanetal03}, \citealt{Farihietal10}). All of this evidence suggests WD\,1337+705 may be accreting circumstellar material. However, no gas disc emission is seen at WD\,1337+705 Burleigh et al. (2010,2011) and no infrared excess has been found \citep{Mullallyetal07}. Consideration must be given to the fact that infrared excesses are not seen at all stars that exhibit photospheric metals \citep{FarihiJuraZuckerman09}, and therefore the lack of an infrared excess should not be taken as firm evidence for there being no circumstellar debris.

\subsection{WD\,1611$-$084 (REJ\,1614$-$085)}
\label{REJ1614$-$085}

WD\,1611$-$084 is the coolest star (\textit{T}$_{\rm eff}$\,=\,38\,840\,K) to display unambiguous signs of circumstellar material. Photospheric detections of C\,{\sc iv}, N\,{\sc v} and Si\,{\sc iv} give $v$$_{\rm phot}$\,=\,$-$40.77$\pm$3.56\,km\,s$^{-1}$. Circumstellar C\,{\sc iv} and Si\,{\sc iv} is present at an average velocity of $-$66.67$\pm$2.05\,km\,s$^{-1}$, with column densities of (9.77$\pm$1.95)x10$^{12}$\,cm$^{-2}$ and (3.53$\pm$0.71)x10$^{12}$\,cm$^{-2}$, respectively. A photospheric O\,{\sc vi} feature is seen in the \textit{FUSE} spectrum of this DA. As seen by \cite{Bannister03}, the velocities of the absorption features are inconsistent across the high ions. Here, the circumstellar C\,{\sc iv} is shifted by $-$23.55$\pm$3.96\,km\,s$^{-1}$ with respect to the photospheric C\,{\sc iv} component; the circumstellar Si\,{\sc iv} is shifted by $-$30.6$\pm$0.64\,km\,s$^{-1}$. Given the poor match to any detected ISM absorption, the non-photospheric material seen in the spectrum of this star may be located near the star in a circumstellar disc, or be further evidence of the vapourisation of planetesimals.

Two components  are seen in the ISM, giving $v$$_{\rm ISM, pri}$\,=\,$-$34.7$\pm$1.5\,km\,s$^{-1}$ and $v$$_{\rm ISM,sec}$\,=\,$-$13$\pm$3.2\,km\,s$^{-1}$. The projected velocity of the G cloud (which is traversed by the sight line to WD\,1611$-$084) is $-$29.26$\pm$1.12\,km\,s$^{-1}$, and does not compare well with $v$$_{\rm ISM, pri}$,  $v$$_{\rm ISM, sec}$  or  $v$$_{\rm CS}$. 

\subsection{WD\,1738+669 (REJ\,1738+665)}
\label{REJ1738+665}

In the case of WD\,1738+669, the averaged $v$$_{\rm phot}$\,=\,30.49$\pm$0.28\,km\,s$^{-1}$, with C\,{\sc iv}, O\,{\sc v} and Si\,{\sc iv} exhibiting clear non-photospheric absorbing components at $v$$_{\rm CS}$\,=\,$-$18.36$\pm$4.23\,km\,s$^{-1}$, giving $v$$_{\rm CS shift}$\,=\,$-$48.53$\pm$4.49\,km\,s$^{-1}$. The measured column densities are (5.55$\pm$0.61)x10$^{13}$\,cm$^{-2}$, (2.35$\pm$0.29)x10$^{11}$\,cm$^{-2}$ and (3.26$\pm$0.17)x10$^{12}$\,cm$^{-2}$ for the circumstellar C\,{\sc iv}, O\,{\sc v} and Si\,{\sc iv}, consistent with the C\,{\sc iv} and Si\,{\sc iv} column densities of (5.01$\pm$0.48)x10$^{13}$\,cm$^{-2}$ and (5.50$\pm$0.13)x10$^{12}$\,cm$^{-2}$ found by \cite{Dupuisetal09}. A circumstellar component is not seen in the N\,{\sc v} doublet as in \cite{Bannister03}, due to the inability to coadd doublet components using the measuring technique utilised here. The \textit{FUSE} O\,{\sc vi} doublet displays two components at $-$32.7$\pm$3.4\,km\,s$^{-1}$ (the circumstellar component) and 14.9$\pm$1.1\,km\,s$^{-1}$ (the photospheric component). This gives a $v_{\rm OVIshift}$\,=\,$-$47.6$\pm$3.6\,km\,s$^{-1}$, similar to that obtained from the \textit{STIS} measurements. Using the S\,{\sc ii} lines, $v$$_{\rm ISM}$\,=\,$-$20$\pm$0.3\,km\,s$^{-1}$, comparing well with $v$$_{\rm CS}$, but not with the predicted $v$$_{\rm LISM}$ of $-$2.91$\pm$1.37\,km\,s$^{-1}$ (due to the LIC).

\subsection{WD\,2023+246 (Wolf\,1346)}
\label{Wolf1346}

Again, neither high ion nor circumstellar absorption is seen at this star. Like WD\,1337+705, the recent studies by \cite{ChayerDupuis10} and \cite{Dupuisetal10} show that in spite of the inclusion of radiative levitation effects, accretion is ongoing. Burleigh et al. (2010,2011) did not find evidence for a circumstellar gas disc emission and no infrared excess has been observed \citep{Mullallyetal07}. 

\subsection{WD\,2111+498 (GD\,394)}
\label{GD394}

No non-photospheric absorption is seen at this star. The \textit{IUE} data shows no C\,{\sc iv}, N\,{\sc v} or O\,{\sc v} absorption and the \textit{GHRS} data only covers the 1290\,\AA\ - 1325\,\AA\ and 1383\,\AA\ - 1419\,\AA\ ranges. $v$$_{\rm ISM}$\,=\,$-$7.6$\pm$1.3\,km\,s$^{-1}$, and the \textit{GHRS} Si\,{\sc iv} absorption lines give a $v$$_{\rm phot}$\,=\,$-$7.28$\pm$1.42\,km\,s$^{-1}$. WD\,2111+498 is well know to have an over abundance of silicon, when compared to model predictions \citep{Holbergetal97}. Coupled with a periodic  variability in the EUV (\textit{p}\,=\,1.150$\pm$0.003), the inhomogeneous accretion of silicon rich material has been suggested to account for the high silicon abundance \citep{Dupuisetal00}. While \cite{Barstowetal03} provided upper limits to the iron and nickel abundances of this star, an analysis by \cite{Chayeretal00} found a near solar Fe abundance using the Fe\,{\sc iii} lines in the \textit{FUSE} spectrum of WD\,2111+498. Since \cite{Chayeretal95} predicted an extremely sub-solar Fe abundance, an external source of material was again invoked to explain the observed abundance. \cite{Vennesetal06} found that WD\,2111+498 in fact had a higher Fe abundance than both WD\,0232+035 and WD\,0501+527. \cite{SchuhDreizlerWolff02} could not model WD\,2111+498 using their self-consistent diffusion/radiative levitation models, citing the accretion of circumstellar material disturbing the diffusion/radiative balance. Since no radial velocity variations have been seen in WD\,2111+498 \citep{Safferetal98}, there is no evidence for a binary companion from which material may be being accreted, suggesting the accretion of circumstellar material may be occurring. However, in a search for circumstellar gas discs, Burleigh et al. (2010,2011) found no emission from Ca\,{\sc ii}, Fe\,{\sc ii} or Si\,{\sc ii} in the optical spectrum of this star, and no infrared excess is detected \citep{Mullallyetal07}.

\subsection{WD\,2152$-$548 (REJ\,2156$-$546)}
\label{REJ2156-546}

The $v$$_{\rm phot}$ of WD\,2152$-$548 is measured at $-$14.94$\pm$0.46\,km\,s$^{-1}$, while $v$$_{\rm ISM}$\,=\,$-$9.2$\pm$0.53\,km\,s$^{-1}$. The projected $v$$_{\rm LISM}$ due to the LIC is $-$9.73$\pm$1.31\,km\,s$^{-1}$, matching $v$$_{\rm ISM}$. \cite{Bannister03} found a hint of a circumstellar component at -1.65$\pm$0.76\,km\,s$^{-1}$ in the coadd of the C\,{\sc iv} doublet. Since this feature was on the edge of detectability in the coadd, and that coaddition in velocity space cannot be preformed here, this possible circumstellar feature is not modelled.

\subsection{WD\,2211$-$495 (REJ\,2214$-$492)}
\label{REJ2214-492}

No circumstellar material is seen at this star. A value of $v$$_{\rm phot}$\,=\,32.33$\pm$1.37\,km\,s$^{-1}$ is obtained from the C\,{\sc iv}, N\,{\sc v}, O\,{\sc v}, Si\,{\sc iv} absorption features, while the S\,{\sc ii} line gives $v$$_{\rm ISM}$\,=\,$-$1.1$\pm$0.4\,km\,s$^{-1}$.

Like the fits of WD\,0948+534 presented earlier, the 1548\,\AA\ component of the C\,{\sc iv} doublet is slightly asymmetric and statistically better fit with one component. A substantial difference in the centroid velocities of the C\,{\sc iv} doublet is noticed both here, with $v$$_{1548\AA}$\,=\,30.5$\pm$0.7\,km\,s$^{-1}$ and $v$$_{1550\AA}$\,=\,37.9$\pm$0.8\,km\,s$^{-1}$, and by \cite{Bannister03}, implying an unresolved circumstellar component may be present. No other lines display a hint of a secondary component.

\subsection{WD\,2218+706}
\label{WD2218+706}

WD\,2218+706 has previously had circumstellar C\,{\sc iv} and Si\,{\sc iv} identified in its UV spectrum \citep{Bannister03}. Here, $v$$_{\rm phot}$ and $v$$_{\rm CS}$ were found at $-$40.04$\pm$1.11\,km\,s$^{-1}$ and $-$17.8$\pm$1.05\,km\,s$^{-1}$, with $v$$_{\rm ISM, pri}$\,=\,$-$15.3$\pm$2.64\,km\,s$^{-1}$ and $v$$_{\rm ISM, sec}$\,=\,$-$1.2$\pm$4.01\,km\,s$^{-1}$. \textit{N}(C\,{\sc iv})\,=\,(1.19$\pm$0.17)x10$^{13}$\,cm$^{-2}$ and \textit{N}(Si\,{\sc iv})\,=\,(7.98$\pm$0.59)x10$^{12}$\,cm$^{-2}$. The line of sight to WD\,2218+706 traverses the LIC, which has a projected velocity of 4.47$\pm$ 1.37\,km\,s$^{-1}$, comparing poorly to $v$$_{\rm CS}$ and the primary ISM component. 

\subsection{WD\,2309+105 (GD\,246)}
\label{GD246}

WD\,2309+105 does not display any signs of circumstellar material in its high ion absorption line profiles. The \textit{STIS} [E140M] data yields a $v$$_{\rm phot}$ of $-$13.45$\pm$0.13\,km\,s$^{-1}$, lining up with the value of $-$13.29$\pm$0.25\,km\,s$^{-1}$ found by \cite{Bannister03}. Given the lack of circumstellar material in the C\,{\sc iv}, N\,{\sc v}, O\,{\sc v} and Si\,{\sc iv} lines, and the O\,{\sc vi} non-detection reported by \cite{Barstowetal10}, the \textit{FUSE} spectrum is not examined here.

\subsection{WD\,2331$-$475 (REJ\,2334$-$471)}
\label{REJ2334$-$471}

Circumstellar material is not observed. A double component fit was statistically preferred for the Si\,{\sc iv} doublet (at 34.00\,km\,s$^{-1}$ and 54.64\,km\,s$^{-1}$) and the N\,{\sc v} 1243\,\AA\ component (at 19.7\,km\,s$^{-1}$ and 43.51\,km\,s$^{-1}$) by \cite{Bannister03}. However, since neither of the components were consistent across the N\,{\sc v} and Si\,{\sc iv} features and the secondary components were not unambiguous, circumstellar material is not modelled here. The $v$$_{\rm phot}$ measured is 38.88$\pm$0.72\,km\,s$^{-1}$, and $v$$_{\rm ISM}$\,=\,14.3$\pm$0.7\,km\,s$^{-1}$.

%%%%%%%%%%%%%%%%%%%%%%%%%%%%%%%%%%%%%%%%%%%%%%%%%%%%%%%%%%%%%%%%%%%%%%%%%%%%%%
%% Acknowledgements                                                         %%
%%%%%%%%%%%%%%%%%%%%%%%%%%%%%%%%%%%%%%%%%%%%%%%%%%%%%%%%%%%%%%%%%%%%%%%%%%%%%%

\section*{Acknowledgements}
\label{acknowledgements}
N.J.D., M.A.B., M.B. and J.F. acknowledge the support of STFC. B.Y.W. would like to acknowledge Guaranteed Time Observer funding for this research through NASA Goddard Space Flight Center grant 005118. We thank Ian Crawford for useful comments and suggestions.

%%%%%%%%%%%%%%%%%%%%%%%%%%%%%%%%%%%%%%%%%%%%%%%%%%%%%%%%%%%%%%%%%%%%%%%%%%%%%%
%% Bibliography                                                             %%
%%%%%%%%%%%%%%%%%%%%%%%%%%%%%%%%%%%%%%%%%%%%%%%%%%%%%%%%%%%%%%%%%%%%%%%%%%%%%%

\end{document}